\documentclass[12pt]{iopart}

\usepackage{color}
\usepackage[utf8]{inputenc}
\usepackage[english]{babel}
\usepackage{graphicx}
\usepackage{hyperref}
\usepackage{epstopdf}
\usepackage{amsfonts}
\usepackage{color}

\newcommand{\ie}{i.\,e.\ }

\usepackage{graphicx}
\usepackage{graphicx}
\usepackage{graphicx}
\begin{document}

\title[Probing and characterizing the growth of a crystal of ultracold bosons and light]{Probing and characterizing the growth of a crystal of ultracold bosons and light}
\author{S. Ostermann$^1$, F. Piazza$^2$, H. Ritsch$^1$}
\address{$^1$Institut f\"ur Theoretische Physik, Universit\"at Innsbruck, Technikerstraße 21, A-6020 Innsbruck, Austria\\
$^2$Max-Planck Institute for the Physics of Complex Systems, Nöthnitzer Straße 38, 01187 Dresden, Germany }
\ead{stefan.ostermann@uibk.ac.at}

\begin{abstract}
The non-linear coupled particle light dynamics of an ultracold gas in the field of two independent counter-propagating laser beams can lead to the dynamical formation of a self-ordered lattice structure as presented in Phys. Rev. X 6, 021026 (2016). Here we present new numerical studies on experimentally observable signatures to monitor the growth and properties of such a crystal in real time. While, at least theoretically, optimal non-destructive observation of the growth dynamics and the hallmarks of the crystalline phase can be performed by analyzing the scattered light, monitoring the evolution of the particle's momentum distribution via time-of-flight probing is an experimentally more accessible choice. In this work we show that both approaches allow to unambiguously distinguish the crystal from independent collective scattering as it occurs in matter wave super-radiance. As a clear crystallization signature we identify spatial locking between the two emerging standing laser waves together creating the crystal potential. For sufficiently large systems the system allows reversible adiabatic ramping into the crystalline phase as an alternative to a quench across the phase transition and growth from fluctuations.  
\end{abstract}

\maketitle

\section{Introduction}

A dilute cold atomic gas illuminated by a laser far off any optical resonance experiences an optical potential with the particles drawn to high or low intensity regions depending on the sign of their linear polarizability. The induced forces on the particles are accompanied by a forward scattering phase shift of the laser beams as well as coherent Rayleigh scattering of light. In the ultracold regime of a BEC, atoms and photons are completely delocalized and the interaction is a collective process from the photon as well as from the atom point of view~\cite{andrews1996direct,inouye1999superradiant,piovella2001quantum}.  It turns out that for a cold and large enough sample of atoms the coupled effective mean field equations for the bosonic atoms and the light are well suited to correctly capture the essence of this complex collective non-linear dynamics~\cite{ostermann2016spontaneous, mueller2016semi}. In particular the Gross-Pitaevski (GP) equation for the atomic quantum gas and the Helmholtz equation for the photons can be used to efficiently describe the coupled nonlinear time evolution in the system. Within this approximation the atoms simply form a dynamic refractive index for the light proportional to their density, while the light creates a dynamic optical potential proportional to the local light intensity guiding the particles. Note that these mean field descriptions of atoms and fields automatically account for collective enhancement of coherent atomic scattering (Bose enhancement), which has been studied in depth in a series of experiments and theoretical models for of single laser excitation from one side. The same is true for stimulated light scattering. As has been known for quite a while now, even for single side illumination by strong enough laser light the system has an instability leading to growing density fluctuations, acceleration and heating~\cite{piovella2001quantum,schneble2003onset,piovella2001superradiant, slama2007superradiant,muller2016semi}. 

In recent work~\cite{ostermann2016spontaneous} we studied a refined setup using a new symmetric and translation invariant geometry of an elongated large cloud of atoms in the field of two counter-propagating independent lasers. To prevent the a priori appearance of an optical lattice, we assume that the two lasers have either orthogonal polarization or sufficiently different frequency to prevent interference and coherent light scattering between the two injected fields. Interestingly, we found that despite the fact that a priori no laser intensity or atomic density modulation is present, above a certain pump intensity threshold, the system spontaneously enters an ordered crystalline phase. In this limit the particles form a periodic density grating and the two laser fields develop into two standing waves, which are coupled and synchronized by the atoms. This can be viewed as crystallization of the particles from homogeneous to periodic order together with the light field. Obviously in this phase transition the system has to choose an effective wave vector and break its continuous translational symmetry. Due to the fact that no mode selective boundaries are present, a continuum of field modes have to be taken into account allowing for the emerging wave-vector, which governs the spatial modulation of the light intensities and the properties of collective excitations in the form of phonon wave packets propagating through the crystal. This distinguishes this crystallization from related cavity based effects as collective atomic recoil lasing~\cite{courtois1994recoil,horak1995recoil, von2004self,slama2007superradiant,kessler2014steering} or transverse self-structuring  by a reflecting mirror where the Talbot length determines the ordering length scale~\cite{robb2015quantum}.

The focus of~\cite{ostermann2016spontaneous} was presenting the basic physical mechanisms leading to this type of crystallization instability. However, many important questions concerning the experimental observability of the crystal phase and the connection to matter wave superradiance were missing in~\cite{ostermann2016spontaneous}. This questions are addressed in the following. Therefore, we present new results to study the onset of crystallization and the characteristics of the buildup of the crystal. We perform detailed simulations and modify our model to contain relevant features for present experimental setups. Thereby we work out common and distinct features to conventional matter wave super-radiance~\cite{schneble2003onset}. One particular goal is of course to identify and study clearly distinguishing experimental signatures of crystal formation. Such signatures can be found from direct time dependent measurements of the scattered or transmitted light on the one hand or by analyzing the particle's momentum distribution via time of flight studies on the other hand.
A extented characterization of properties of the crystalline phase is another central aim of this work. From a practical point of view it seems that many of those  features are not so easy to observe as current experiments are limited in particle number. While in principle for any particle number an instability and ordering threshold exists, the necessary power to reach the ordering instability at sufficiently large detuning is very high and simultaneously leads to heating and fast particle loss. This strongly limits the observation time and accuracy of the scattered field measurements. In order get some estimates of these perturbations we phenomenologically introduce particle loss in our model equation~\cite{ostermann2016spontaneous} so that its effect on the real time dynamics can be predicted.

In particular during the initial time evolution of the system, where only weak light back scattering occurs, the fields act rather independently and the predicted signals and atomic momentum distributions are very similar to matter wave superradiance~\cite{schneble2003onset, inouye1999superradiant,mueller2016semi}. One possible signature includes a careful analysis of the instability threshold of a homogeneous density for single side versus symmetric pumping. In this respect we perform very detailed studies of the the short time dynamics, varying the pump geometry from symmetric pumping for ideal crystallization via asymmetric pumping towards the single side pump case as the standard configuration where matter wave super-radiance is usually observed. Such detailed predictions should be valuable to distinguish the processes even at current experimental limitations. Indeed the onset of the predicted crystal phase could be observed in a recent experiment which implements the proposed geometry~\cite{dimitrova2017observation}.

As the atom-field crystal corresponds to an at least metastable ground state of the system one aim is certainly to prepare it as good as possible. Since a sudden quench leaves much entropy in the final crystal, an alternative approach is the controlled adiabatic ramp into the crystalline phase. Varying the switch time from a sudden quench across threshold to slow almost adiabatic ramp, we can address the created entropy via the reversibility of the ordering dynamics. Of course a very slow ramp increases the effect of background heating which again increases the final entropy in the crystal.  In fact, the best way to reach a clean crystal would be using a time dependent laser field found by optimal control algorithms~\cite{doria2011optimal,bucker2013vibrational}. This is certainly interesting but goes beyond the scope of this work.

\section{Model}\label{sec:model}
As basis of our analysis we use the effective mean field model introduced in our previous work in~\cite{ostermann2016spontaneous}. Here we will only briefly describe its main properties to give a self contained basis for our following detailed studies. For the detailed description and derivation we refer to section two in~\cite{ostermann2016spontaneous}.

\begin{figure}
\centering
\includegraphics[width=0.75\textwidth]{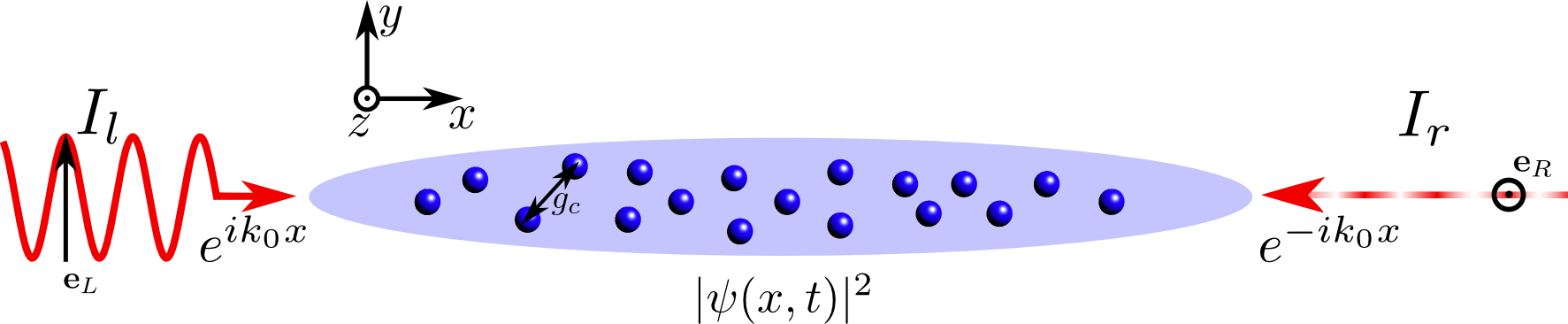}
\caption{Schematic view of the setup. An elongated BEC is interacting with two counterpropagating plane waves of orthogonal polarization. The beams are assumed to be far off-resonant from the atomic transition to avoid polarization mixing.}
\label{fig:setup}
\end{figure}

An extended atomic BEC interacting with two off-resonant and counter-propagating independent electromagnetic plane waves injected from two opposite sides (see~\fref{fig:setup}) is studied.  The two electromagnetic fields Stark shift the atomic ground state and create an optical potential for the BEC atoms, which in turn modify the field propagation as they constitute an effective density dependent index of refraction.
The dynamics of this setup can be efficiently described by a coupled system of mean field equations. The BEC dynamics is well approximated by the Gross-Pitaevski (GP) meanfield regime, where the condensate wavefunction $\psi(x,t)$ satisfies the equation:
\begin{equation}
i \hbar\frac{\partial}{\partial t}\psi(x,t)=\left[\frac{-\hbar^2}{2m}\frac{\partial^2}{\partial x^2}+V_\mathrm{opt}(x)\right]\psi(x,t)+\frac{g_cN}{A}|\psi(x,t)|^2\psi(x,t).
\label{eqn:GPE}
\end{equation}
Here $m$ denotes the particle mass, $g_c$ is the effective atom-atom interaction strength, $N$ denotes the atom number, $A$ is the transversal BEC cross section and $V_\mathrm{opt}(x)$ stands for the optical potential generated by the electromagnetic fields. The treatment of the dynamics is restricted to one dimension, i.e. we assume the dynamics in the $y$ and $z$ directions is frozen due to the transverse confinement of the BEC by an external trapping potential. Actually the Gaussian pump laser beams could in principle provide such a potential, but we refrain from this complication in our model  and assume laser beams much broader than the BEC.  Such an one dimensional treatment is only valid if the BEC's chemical potential $\mu$ is much smaller than the characteristic trap frequency $\mu<<\hbar \omega_{y,z}$.

Within this limit the optical potential, which determines the BEC dynamics in $x$ direction can be written as
\begin{equation}
V_\mathrm{opt}(x)=-\alpha\left[|E_L(x)|^2+|E_R(x)|^2\right]+V_{ext}
\end{equation}
where $\alpha$ denotes the atomic polarizability.
The first term corresponds to an optical dipole potential in the low-saturation regime where $E_{L,R}(x)$ are the two field envelopes of the electromagnetic fields $\mathbf{E}_{L,R}(x,t)=\left[E_{L,R}(x)e^{i\omega_Lt}+c.c.\right]\mathbf{e}_{L,R}$ (with $\mathbf{e}_L\cdot\mathbf{e}_R=0$) which satisfy the Helmholtz equations
\begin{equation}
\frac{\partial^2}{\partial x^2}E_\mathrm{L,R}(x)+k_0^2 (1+\chi(x))E_\mathrm{L,R}(x)=0
\label{eqn:Helmh_total}
\end{equation}
with the wavenumber $k_0$ of the incoming beams and the BEC's susceptibility $\chi(x)$. The susceptibility depends again on the condensate's density via
\begin{equation}
\chi(x)=\frac{\alpha N}{\epsilon_0A}|\psi(x)|^2,
\label{eqn:susc}
\end{equation}
where $\psi(x,t)$ is the solution of~\eref{eqn:GPE} and $\varepsilon_0$ denotes the free space permittivity. $V_{ext}$ is an externally prescribed potential chosen as a square box potential for computational convenience.   

Again, the constant $\alpha$ in Eqs.~\eref{eqn:Helmh_total} and~\eref{eqn:susc} denotes the absolute value of the real part of the atomic polarizability. Hence, the absolute minus sign in~\eref{eqn:Helmh_total} indicates (far off-resonant) red-detuning from any atomic transition, \ie high field seeking atoms. In addition, this impies that the potential depth is proportional to $|\Gamma/\delta|$ where $\Gamma$ is the spontaneous decay rate and $\delta$ the detuning relative to the internal atomic transition. As a result, the potential depth scales with $|\delta|^{-1}$. In the following we will use the dimensionless quantity
\begin{equation}
\zeta:=\frac{\alpha N}{\varepsilon_0\lambda_0 A}=\frac{\alpha}{\varepsilon_0}n\frac{L}{\lambda_0}
\end{equation}
to quantify the light matter interaction where $\lambda_0$ is the wavelength  of the incoming laser beams and the three-dimensional density $n$ of the homogeneous BEC which is supposed to have length $L$ in $x$ direction.

The system described by the set of equations~\eref{eqn:GPE}-\eref{eqn:susc} is, apart from $V_{ext}$, a priori translation invariant. If one simulates the dynamics described by this set of equations in a self-consistent manner, \ie updating the electric field distribution for the corresponding BEC density within each timestep of the GPE solution, one finds the following remarkable behaviour. As long as the BEC density is homogeneous the light fields propagate through the sample without any spatial modulation. However, we have shown that when the pump light intensity exceeds a certain critical value, the light spontaneously crystallizes together with the atoms by breaking the translation invariance. For a detailed description of the numerical methods used in this context we refer to Appendix B of~\cite{ostermann2016spontaneous}.

\section{Real-time observation of spontaneous crystallization via the back-scattered light fields}

\label{sec:crystallization}
When the driving intensity exceeds a certain critical value, small fluctuations will grow and break the translation invariance. Performing a linear stability analysis of eqs.~\eref{eqn:GPE}-\eref{eqn:susc} as it is presented in~\cite{ostermann2016spontaneous} leads to a critical intensity value (in the limit $L\rightarrow\infty$ and $N={\rm const.}$) which depends inversely on the particle number in the form:
\begin{equation}
I_{c}^{\rm L,R}=\frac{c E_{rec}}{\lambda_0A}N\frac{1}{\zeta^2}=cE_{rec}\frac{\varepsilon_0^2}{\alpha^2 }\frac{1}{n}\frac{\lambda_0}{L}.
\label{eqn:Icrit}
\end{equation}
Here we introduced the recoil energy $E_\mathrm{rec}:=\hbar^2 k_0^2/(2m)$. In the following we will use the quantity $I_0:=c E_{rec}/(\lambda_0 A)$ as the natural unit for intensities.

We see that the threshold~\eref{eqn:Icrit} scales with the inverse square of the atomic polarizability $\alpha^{-2}\propto\delta^2$ and hence grows approximately with the square of the laser detuning from resonance.
Qualitatively this behavior can be traced back to the non-linear nature of the interactions of the considered system. Initially, for the homogeneous system, the backscattered field amplitude is proportional to $\alpha$. This coherently backscattered field interferes with the incoming laser leading to an initial potential modulation proportional to $\alpha$, which then translates into an $\alpha^2$ term originating from the non-linearity of the refractive index term in the Helmholtz equation.
Note that also the heating via spontaneous scattering exhibits the same detuning scaling the essential route to small heating at threshold is a large particle number. 

\begin{figure}
\centering
a)\includegraphics[width=0.4\textwidth]{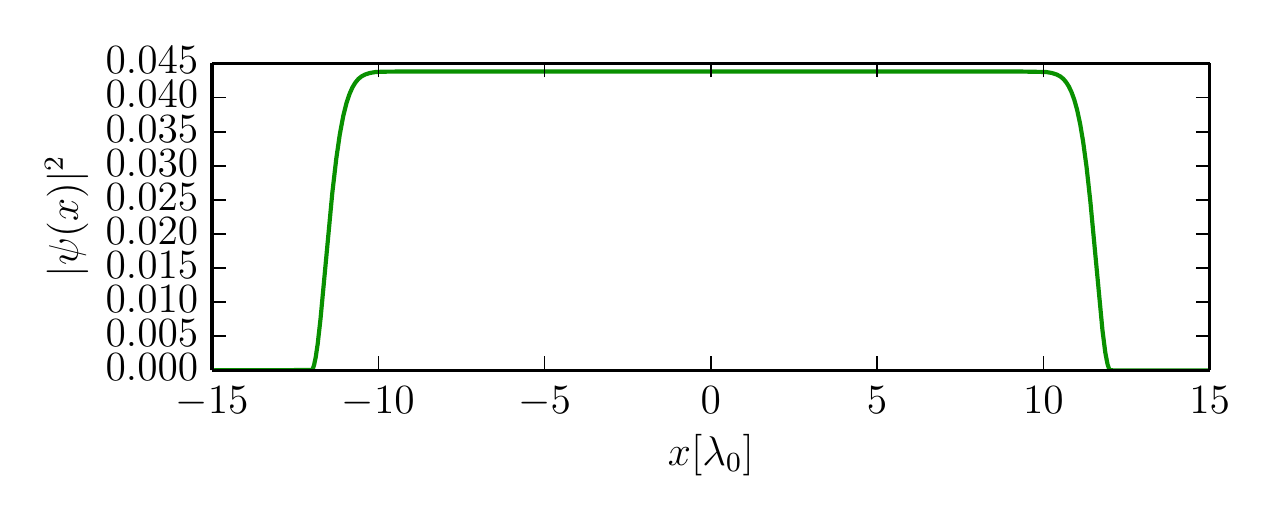}
c)\includegraphics[width=0.4\textwidth]{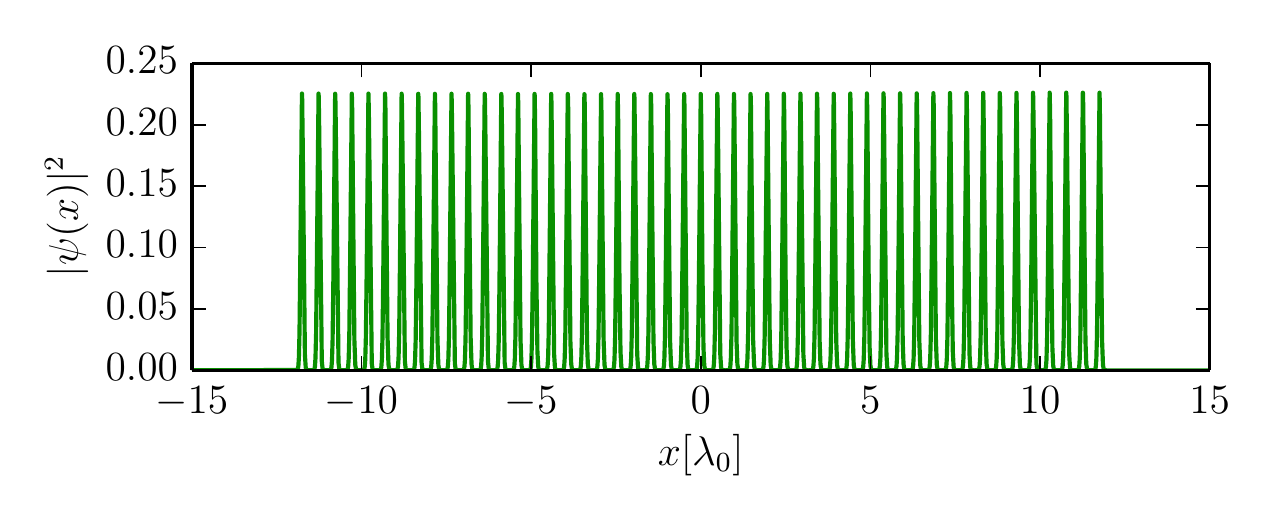}\\
b)\includegraphics[width=0.4\textwidth]{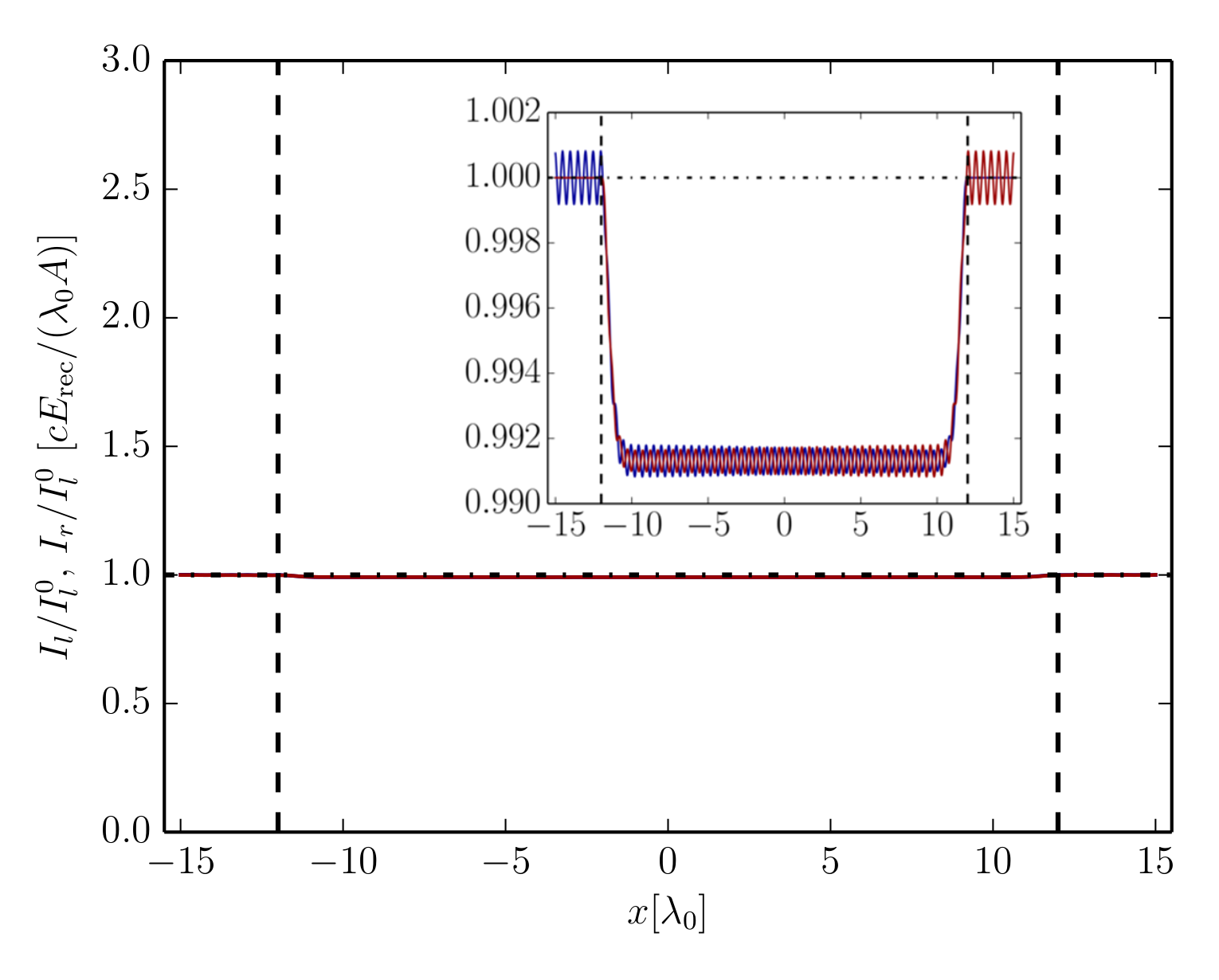}
d)\includegraphics[width=0.4\textwidth]{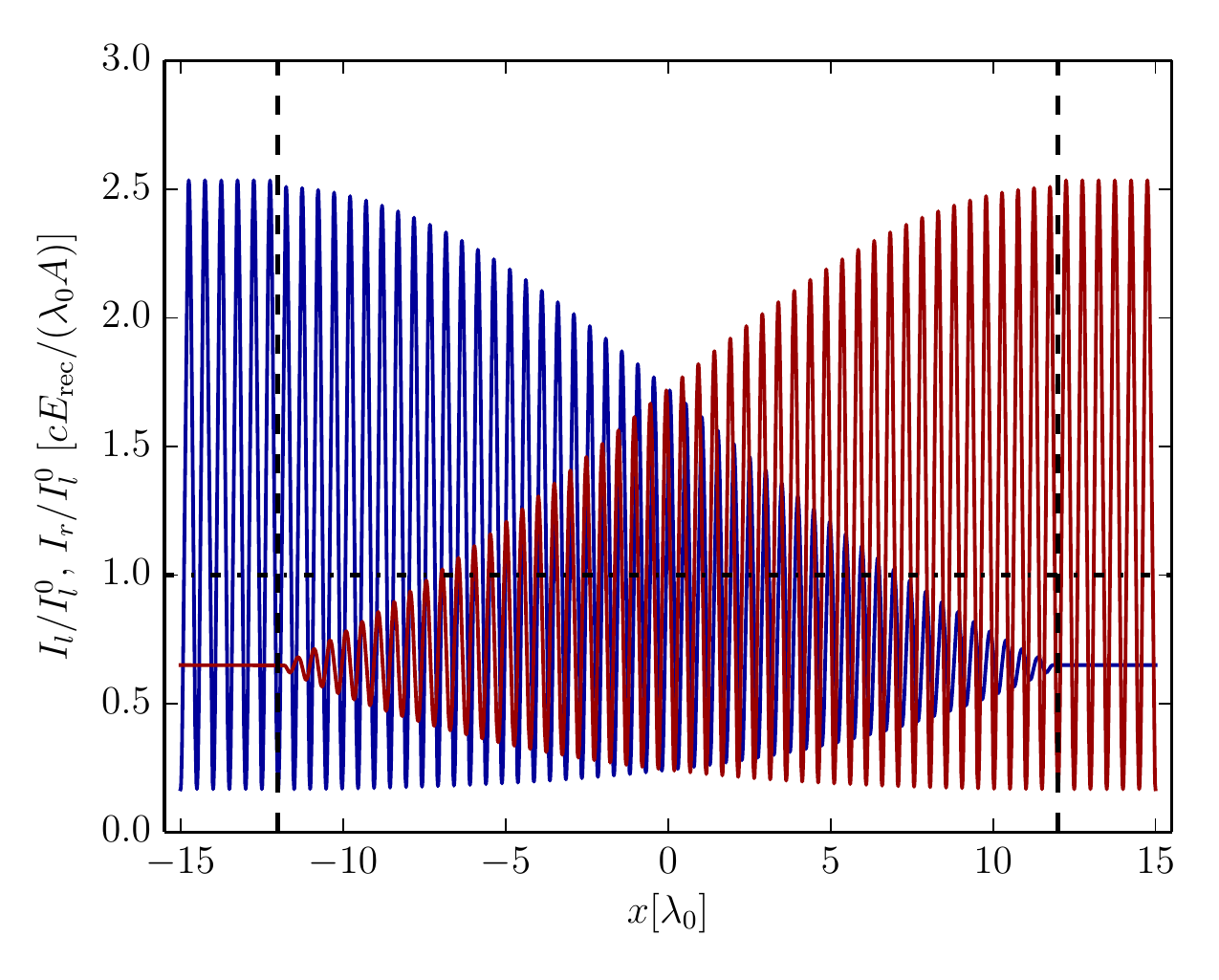}
\caption{Steady state atomic density and light intensity as function of position along the axis for effective coupling strength $\zeta=0.2$ and a calculation box length of $L=30 \lambda_0$ which is significantly larger than the external deep trapping potential wells starting at $x=-12\lambda_0$ and ending at $x=12\lambda_0$. The intensity threshold for this regime lies at $I_c^L=I_c^R=12.5 I_0$ (see also~\fref{fig:rIdep}). Figure a) shows the particle density and b) the light intensity below threshold for an incoming light intensity $I_{l,r}^0=10 I_0$. The y-axis is rescaled with respect to this incoming light intensity. The inlay zooms to values close to $I_l/I_l^0=I_r/I_r^0=1.0$ showing that only a small part of the light is reflected. Part c) and d) show the same densities for $I_{l,r}^0=200 I_0$, \ie far above threshold where about half of the incoming intensity is back reflected by the self formed lattice.}
\label{fig:density}
\end{figure}

\begin{figure}
\centering
a)\includegraphics[width=0.3\textwidth]{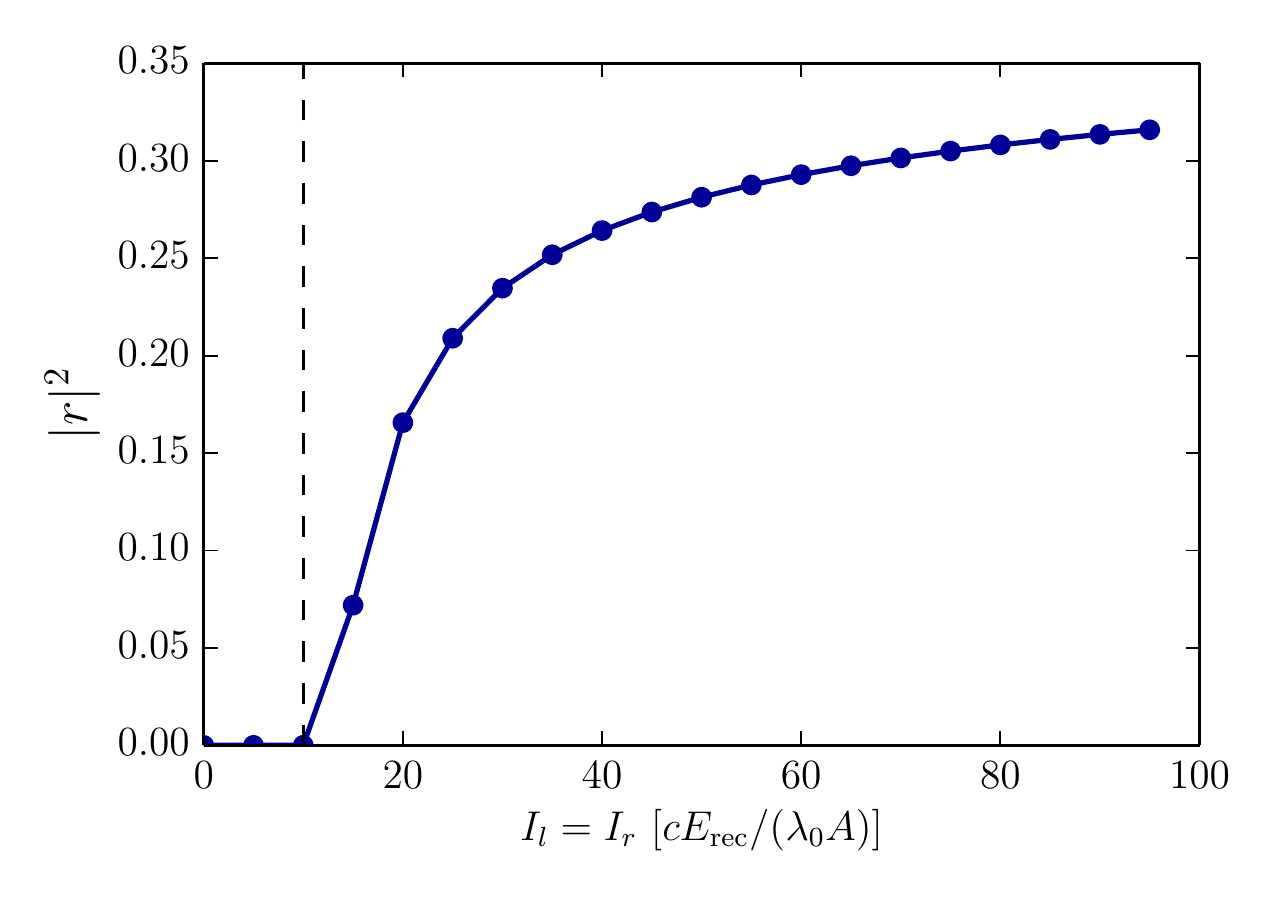}
\hfill
b)\includegraphics[width=0.3\textwidth]{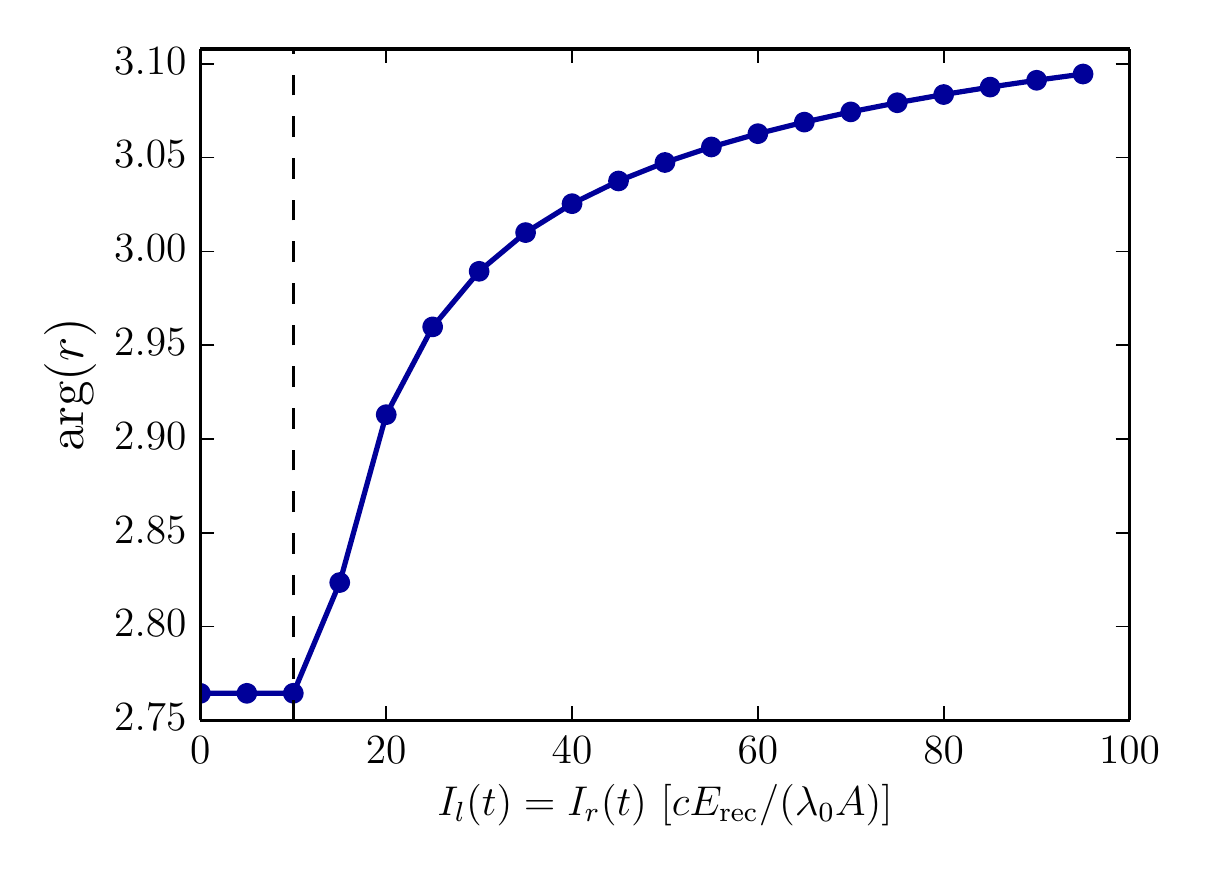}
\hfill
c)\includegraphics[width=0.3\textwidth]{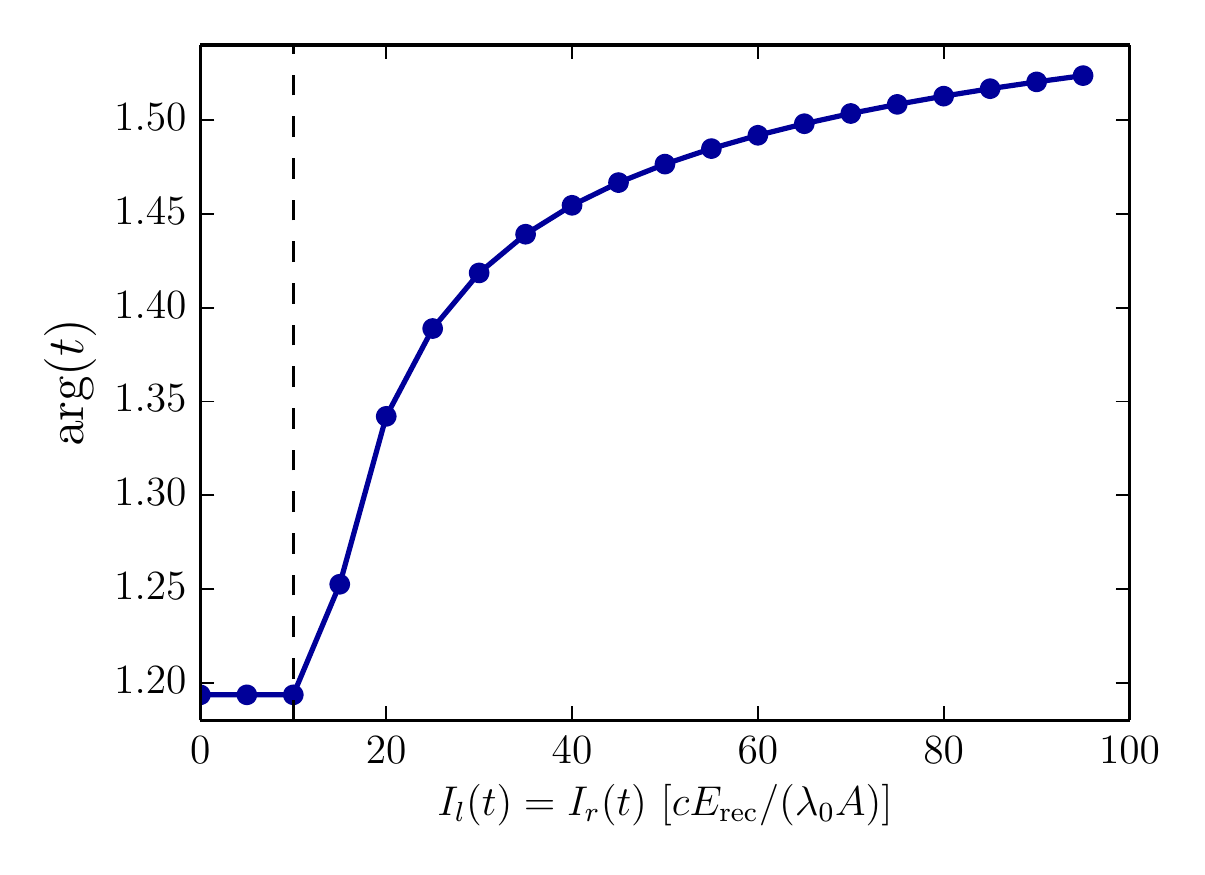}
\caption{Dependence of the total reflection coefficient $r$ on the pump intensity: a) shows the absolute value squared, b) the phase of the reflection coefficient c) the phase of the transmission coefficient for $L=100\lambda_0$  and all other parameters are chosen as in~\fref{fig:density}).}
\label{fig:rIdep}
\end{figure}

\begin{figure}
\centering
\includegraphics[width=0.4\textwidth]{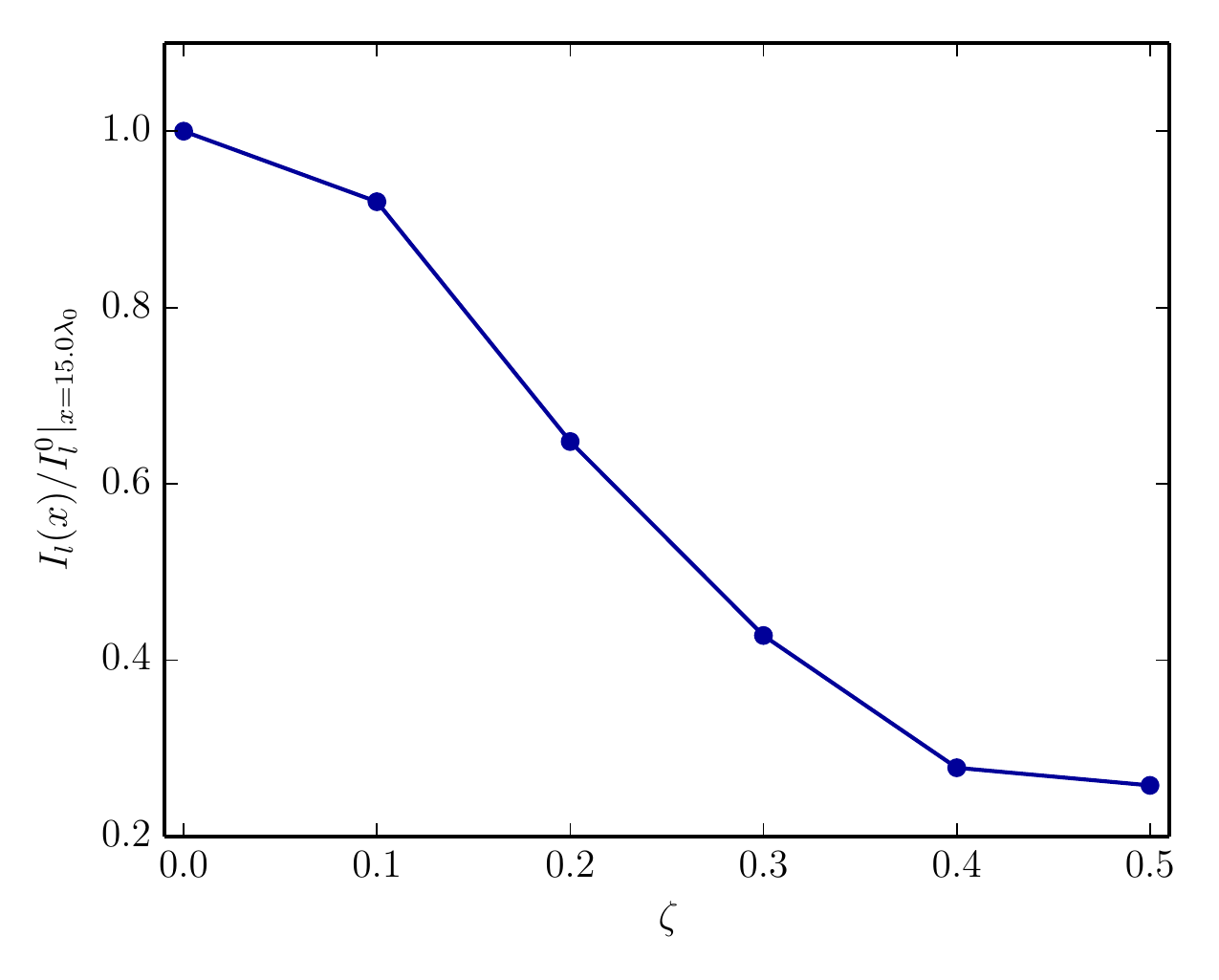}
\caption{Relative fraction of the intensity transmitted through the whole atomic crystal as shown in~\fref{fig:density}d) for different values of the effective coupling parameter $\zeta$ with all other parameters are chosen as in~\fref{fig:density}d).}
\label{fig:Ibound_zdep}
\end{figure}

For intensities above threshold a periodic modulation of the optical potential for the atoms is formed spontaneously and the atoms crystallize into a periodic structure together with the light fields. Since we consider the full electromagnetic field in our model (see eq.~\eref{eqn:Helmh_total}) all possible modes can be addressed in general. Therefore, the lattice spacing is emergent and the the new solutions of~\eref{eqn:Helmh_total} are plane waves of the form $E_\mathrm{L,R}=C \exp(\pm i k_\mathrm{eff}x)$ propagating with a modified wavenumber
\begin{equation}
\label{eqn:keff}
k_{\rm eff}=\frac{2\pi}{\lambda_0}\sqrt{1+\frac{\alpha}{\epsilon_0}n}
\end{equation}
which is a result of the modified BEC density $n$.
Examplary solutions for the system's ground state of the coupled set of equations~\eref{eqn:GPE}-~\eref{eqn:susc} for the density and the fields are shown in~\fref{fig:density}. A box potential is included in the GPE equation in order to simulate the system boundaries. Obviously, the density stays homogeneous (up to small modulations which are finite size effects) for intensities below threshold and so does the intensity distribution (see~\fref{fig:density}a) and b)).
Above threshold the continuous translational symmetry is broken and a periodic optical potential is generated for the atoms. This spontaneous crystallization relies on the fact that a counter propagating field component with the same polarization is generated from the BEC which acts as a Bragg mirror for the fields. In the limit of deep local traps the behavior resembles very closely the case of an array of mobile beam splitters which can be used as toy model to understand optical lattice dynamics as shown in fig.5 of ref.\cite{ostermann2014scattering}. Hence, we expect the complete system's reflectivity to grow drastically as soon as the intensity of the incoming light fields lies above the threshold value. Indeed, both the absolute value and the phase of the reflection and transmission coefficient show this expected behavior (see~\fref{fig:rIdep}). The change of the phase of the transmission spectrum at threshold also implies that the threshold could also be measured by phase-contrast imaging of the BEC~\cite{andrews1996direct,szigeti2009continuous}.
This process leads to the fact that the field amplitudes outside the system boundaries lie far below the initial incoming laser intensity in~\fref{fig:density}d). In this case it is of specific interest if there is a certain upper boundary value for the index of refraction. In~\fref{fig:Ibound_zdep} we plot the value of the constant intensity part outside the BEC for different light interaction strengths $\zeta$. Obviously, this value decreases drastically with increasing $\zeta$. However, due to the interaction of the two light fields via the BEC atoms the intensity pattern always adjusts in a way that the envelope of the total intensity is not modulated at all. This "phase-locking" of the intensity patterns will be studied in more detail below.

The system's properties described above lead to the fact that the phase transition from a homogeneous BEC to a spontaneously formed crystal can be non-destructively observed by measuring the light reflected from the atomic cloud. In fact, the whole time-dynamics of the system can be mapped out by measuring these properties of the light fields.

\begin{figure}
a) \includegraphics[width=0.4\textwidth]{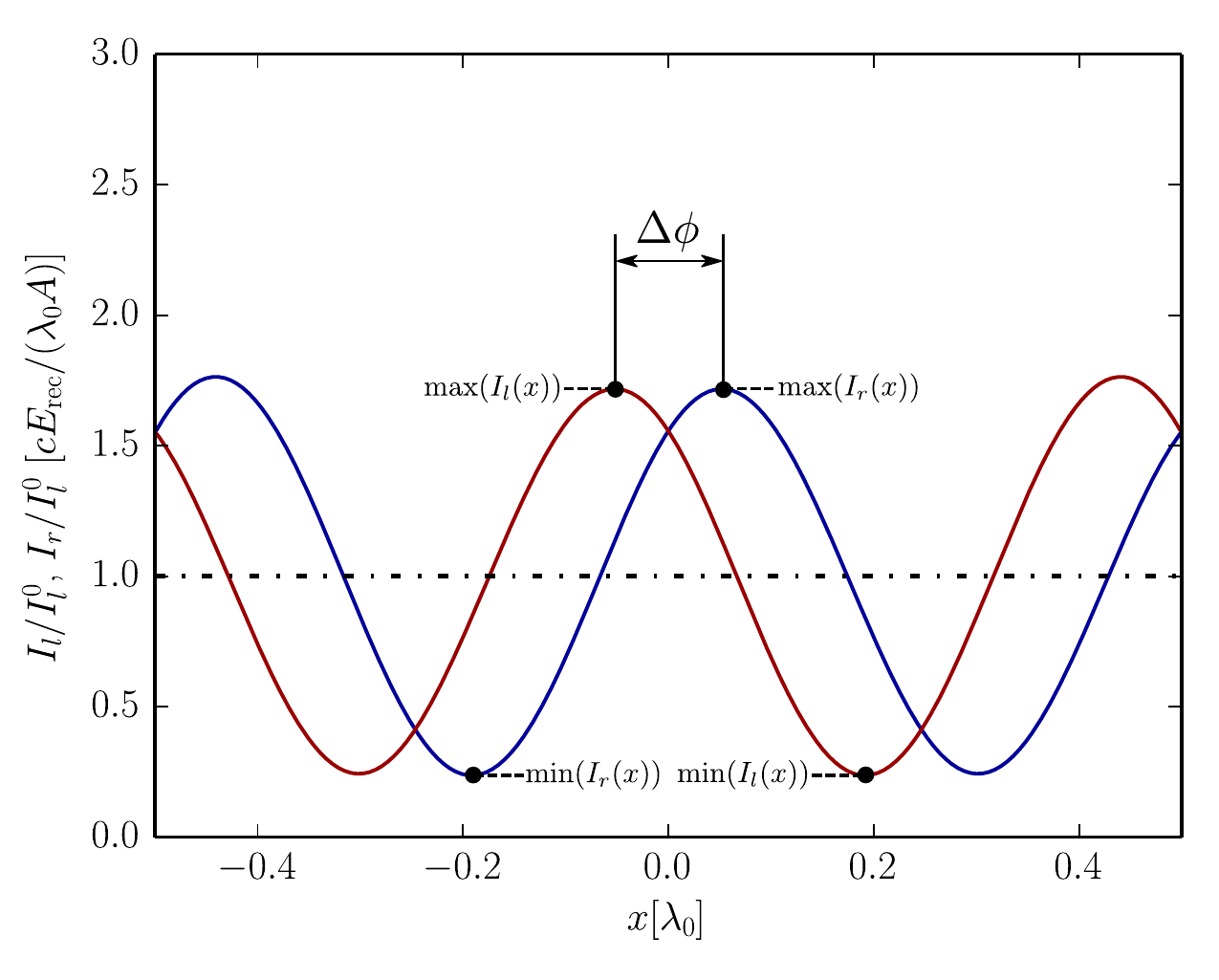}\hspace{1cm}
b) \includegraphics[width=0.4\textwidth]{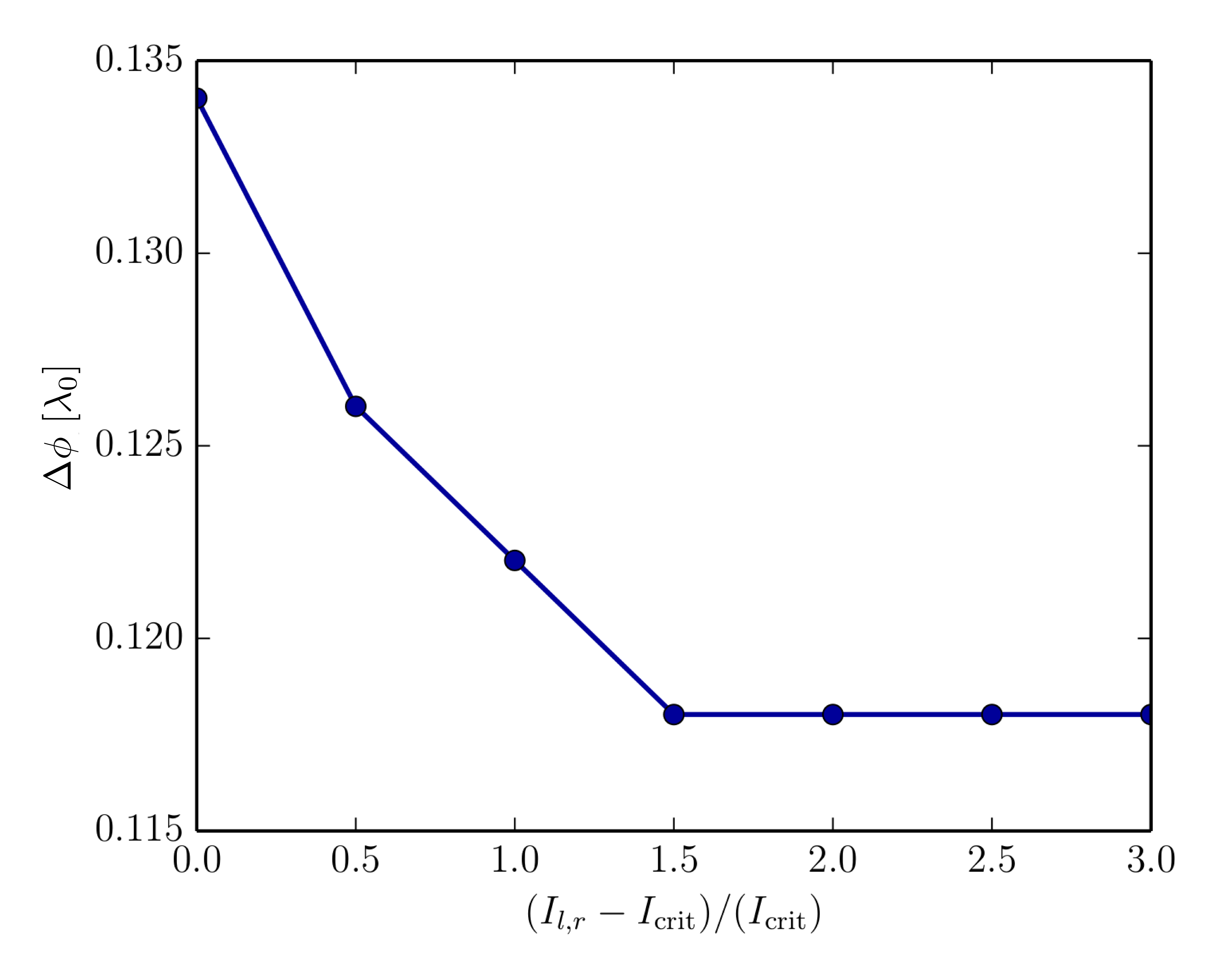}
\centering
\caption{a) Enlarged view of the central part of intensity distribution of the two fields as shown in~\fref{fig:density}d) exhibiting a relative shift of the intensity maxima (phase shift) of the left and right injected field . In figure b) we show this phase shift for the stationary ground state for different incoming laser intensities. }
\label{fig:thres_phaselock}
\end{figure}

Let us now turn to a closer analysis of the light intensity distributions. If one zooms into the region around zero in~\fref{fig:density}d) one finds that the intensity distributions of the field coming from the right side and the field coming from left are not in phase (see~\fref{fig:thres_phaselock}a)). This feature of the solution of the Helmholtz equation~\eref{eqn:Helmh_total} is a fundamental feature of the crystal phase. The fact that the maxima of the single light fields do not coincide leads to a strong coupling of the atoms to the fields due to the large intensity gradient which is felt from each individual light field. This is a fundamental difference to for example standard optical lattices. Interestingly, the dynamics of the full system again leads to a potential which in total results in an overall homogeneous density modulation.
The question one can address in this context is the intensity dependence of the dephasing $\Delta \phi$ as it is defined in~\fref{fig:thres_phaselock}a). As one can see from~\fref{fig:thres_phaselock}b), this dephasing is getting smaller for growing intensities until it reaches a certain constant value.

\section{Time evolution of atomic momenta from Bragg-diffraction}\label{sec:bragg}
Event though the non-destructive measurement of the properties of the coupled BEC-light crystal via the reflected light fields is a special feature of the studied system, it is of specific interest how the crystallization process and its properties can be observed with a standard experimental technique like Time-of-Flight (TOF) measurement.
It consists in releasing the external trapping potential and letting the BEC expand ballistically. After a certain time of flight (typically a few ms) a laser pulse is imposed from the side and the absorption image generated by the BEC is then captured on a CCD camera. If the far-field regime is reached for long enough time evolution, the absorption image corresponds to the BEC momentum distribution.
\begin{figure}
\centering
a) \includegraphics[width=0.3\textwidth]{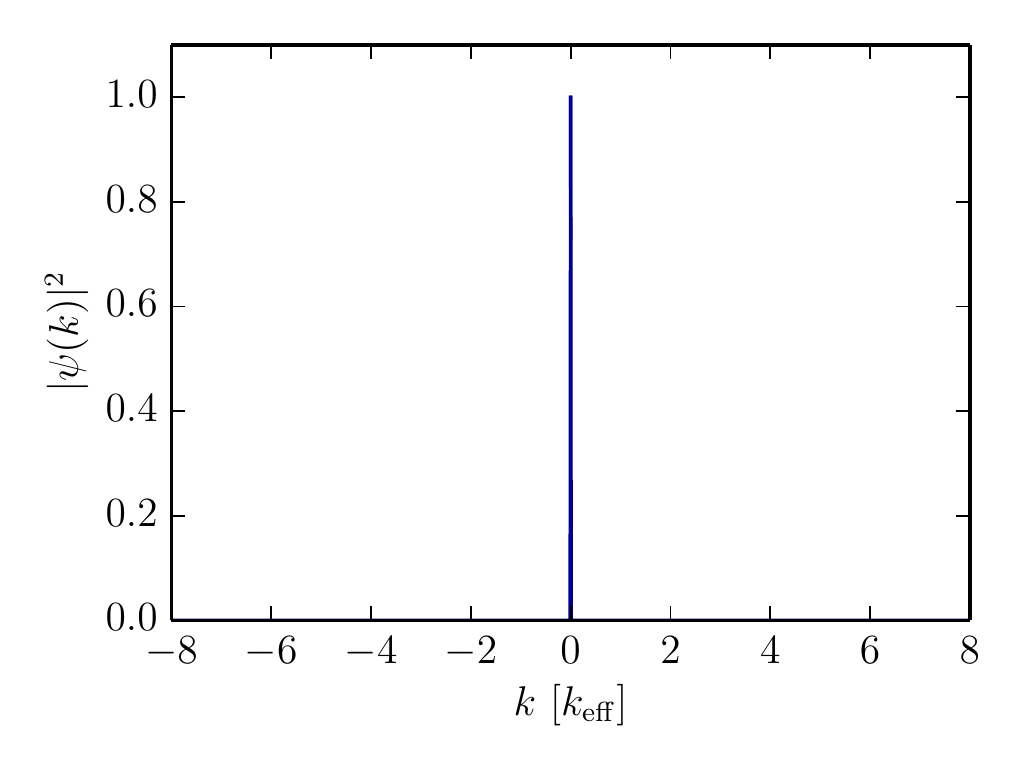}
b) \includegraphics[width=0.29\textwidth]{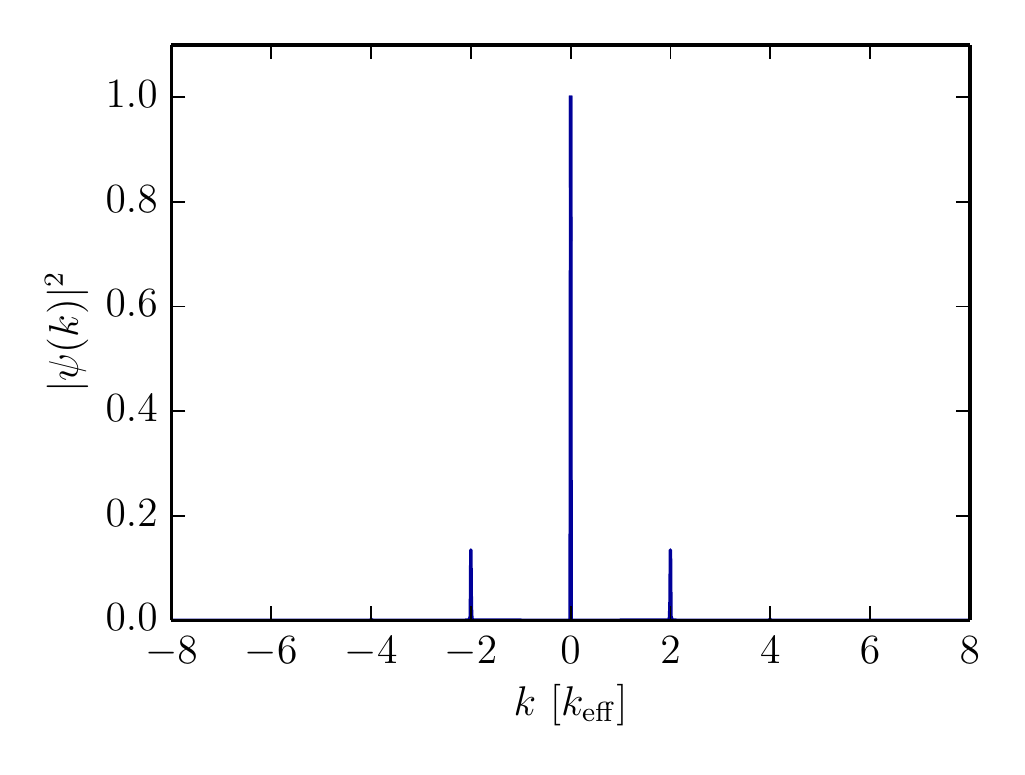}
c) \includegraphics[width=0.29\textwidth]{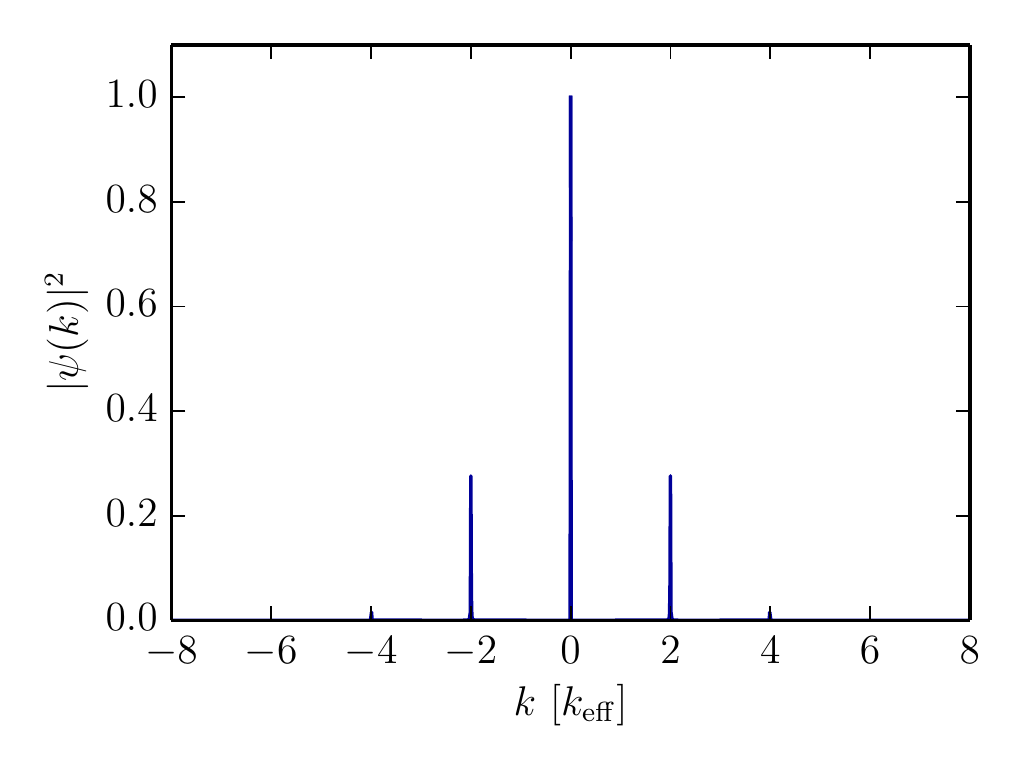}
\caption{Momentum distribution (Bragg diffraction pattern) for the system ground state obtained via for imaginary time evolution for different pump intensities. a) $I=10 I_0$, b) $I=100I_0$ and c) $I=200I_0$. The chosen system parameters are: $\zeta=0.1$, $L=100\lambda_0$, $g_c N=1.0 E_{rec}$ with the height of the $k=0$ peak normalized to 1.}
\label{fig:Bragg_peaks}
\end{figure}
Therefore, the corresponding TOF measurement outcome for a specific solution of Eqs.~\eref{eqn:GPE}-\eref{eqn:Helmh_total} can be easily obtained by calculating the Fourier transform of the BEC wave function. In~\fref{fig:Bragg_peaks} a typical example of such a momentum space distribution at different intensities of the crystallization procedure is shown.

Of course, the homogeneous BEC contains only the $\hbar k=0$ momentum component (see~\fref{fig:Bragg_peaks}a). However, for intensities above threshold the $\pm2\hbar k$ and higher momentum components are getting populated (see~\fref{fig:Bragg_peaks}b-c). This is a clear indication for an emerging periodic structure. If it is possible to verify that no interference is possible between the two counterpropagating light fields this measure can be used as a clear indication for an emerging crystal. As a matter of fact, the $\pm2\hbar k_\mathrm{eff}$ peak dominates among the higher momentum peaks which are present but small.This happens even at strong intensities, indicating that the emerging modulation is mostly sinusoidal.

\begin{figure}
\centering
\includegraphics[width=0.5\textwidth]{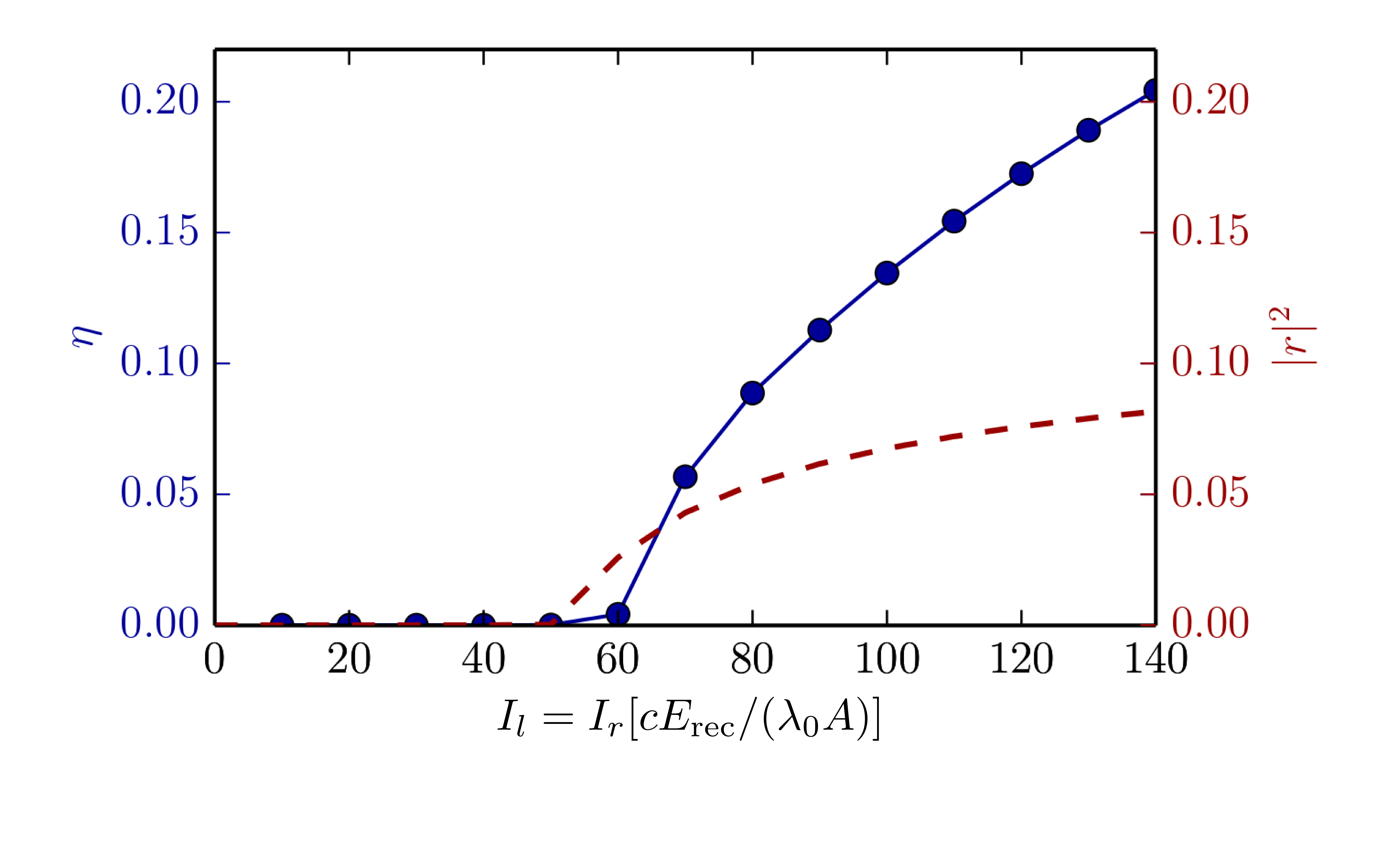}
\caption{Intensity dependence of the ratio $\eta$ of the height of the central peak and versus the $2k_\mathrm{eff}$ peak (solid blue, left axis). One recognizes good agreement of the growth point of this ratio with the threshold estimation via the reflected light (dashed red, right axis). The parameters are the same as in~\fref{fig:Bragg_peaks}.}
\label{fig:Bragg_thres}
\end{figure}

Another important observation is that the height of the $2\hbar k_\mathrm{eff}$ peak relative to the central peak behaves as the system's reflectivity. This means that the value of the ratio
\begin{equation}
\eta=\frac{|\psi(k=2k_{\rm eff})|^2}{|\psi(k=0)|^2}
\label{eqn:eta}
\end{equation}
where $\psi(k)=\frac{1}{L}\int dx e^{ikx}\psi(x)$ is the Fourier transform of the BEC wavefunction can be used to measure the crystallization threshold~\eref{eqn:Icrit}. In this case the TOF measurement has to be performed for several different pump intensities and the ratio $\eta$ has to be estimated for each measurement. The resulting threshold estimation is in good agreement with the one via the system's reflectivity $|r|^2$ (see~\fref{fig:Bragg_thres}).
As a result, the parameter $\eta$ is an easy to access quantity to measure the physical properties of the system directly via the TOF image. 

\section{Scaling of the collective dynamics}\label{sec:partloss}

\begin{figure}
\centering
\includegraphics[width=0.5\textwidth]{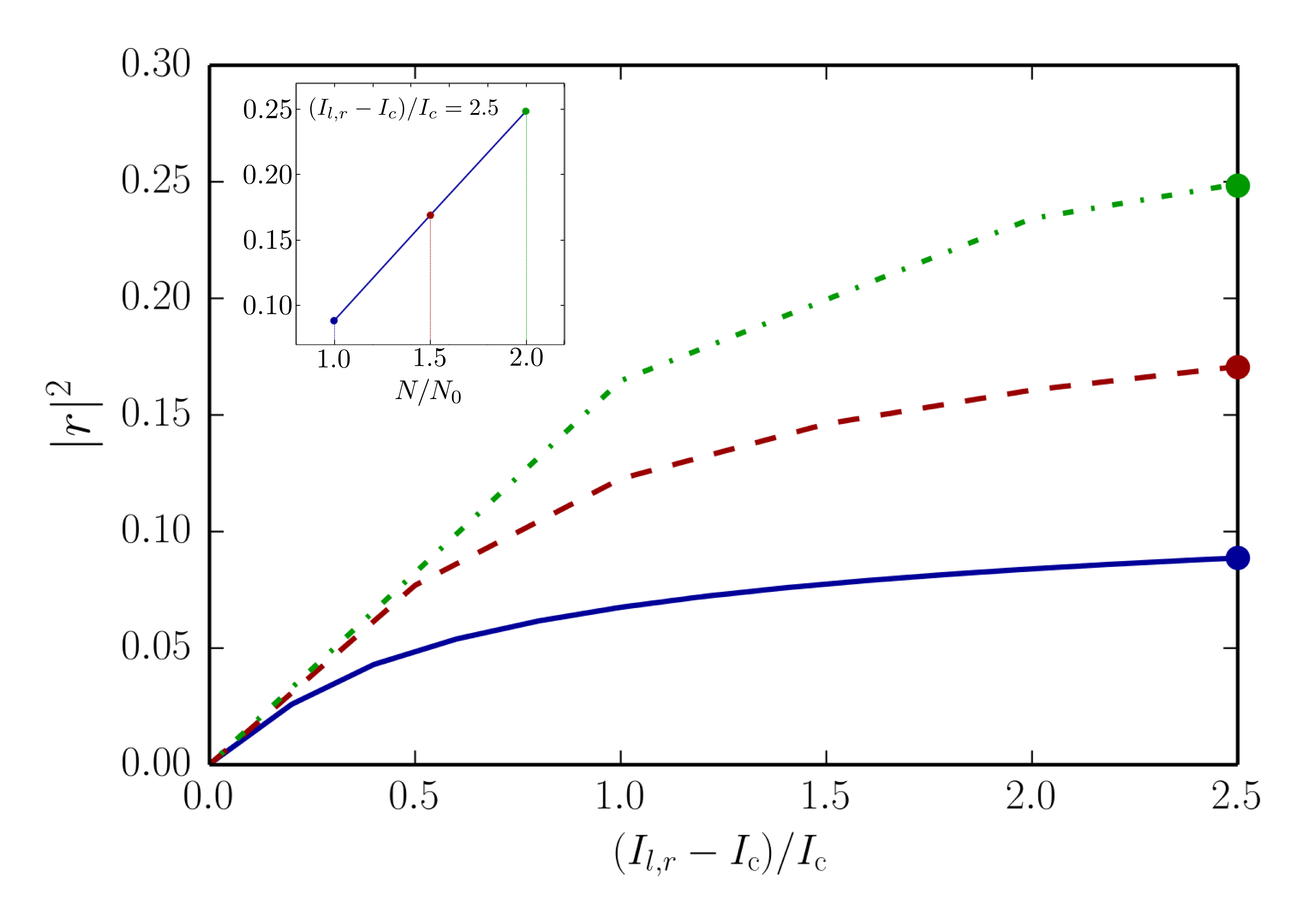}
\caption{Particle number scaling of the reflection coefficient. The particle number of the dashed red curve is a factor of 1.5 larger and the one for the dash dotted green curve is a factor 2.0 bigger than the one of the solid blue curve. Note that the axes have been rescaled by the corresponding critical intensities. ($\zeta=0.1$, $L=100\lambda_0$, $g_c N=1.0 E_{rec}$). The inset shows the linear dependence of the reflectivity on the particle number.}
\label{fig:r_Ndep}
\end{figure}

One centrally important feature of the crystallization process is its collective nature. Since all atoms in the BEC interact synchronously via the scattered light field we expect a strong particle number dependence of the crystallization process. In particular the more particles the less the threshold power and saturation of the atomic transition. The collective nature of the scattering remains also in the ordered phase above threshold, which can be directly verified from the reflectivity for different particle numbers $N$. Indeed, the total system's reflectivity above threshold increases with the particle number (see~\fref{fig:r_Ndep}) until a saturated value is reached where most of the light is reflected.
It is an interesting question here, what one could expect in the thermodynamic infinite particle number limit. Is there an instability when all the light from one side is reflected and the system acts as a perfect mirror. What is the dynamics of the transition region here. For a transversely pumped case we previously found a fractioning of the system\cite{griesser2013light} . Unfortunately the required system sizes are too big for a straightforward analysis here.

For current realistic experimental setups the particle number is limited to well below $10^6$ so that the ordering threshold occurs at rather high saturation parameters. To better understand what is going on in this case particle loss has to be included into the model. This allows how long the coherent collective growth of the reflection signal persists against the reduction of reflection due to particle particle loss. In any case only a transient signal can be expected. Since a full treatment of particle loss via the spontaneous emission heating rates would be very demanding, we simple introduce a phenomenological toy loss mechanisms to study the effect.

\begin{figure}
\centering
\includegraphics[width=0.5\textwidth]{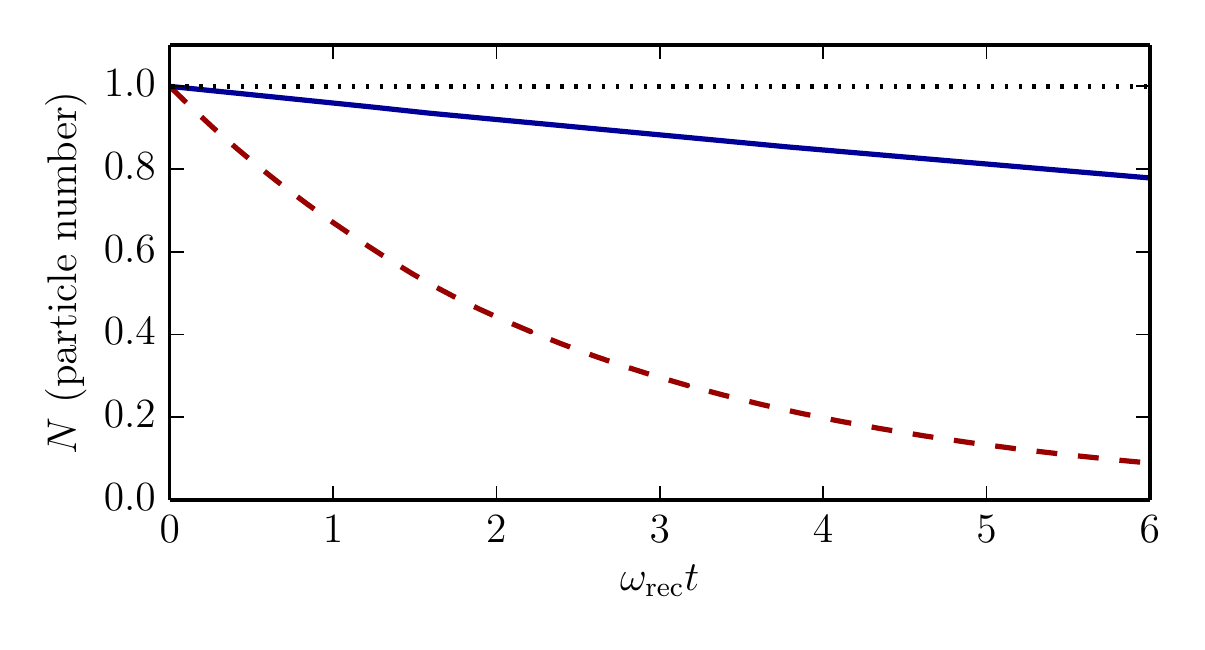}
\caption{Time evolution of the particle number for different choices of the intensity dependent particle loss rate $\gamma$ as defined in~\eref{eqn:part_loss}. $\gamma=0.001$ (solid blue), $\gamma=0.01$ (dashed red). The dotted black line corresponds to $\gamma=0$. Other parameters: $I_l=I_r=100 I_0$, $L=10\lambda_0$, $g_cN=1.0 E_{rec}$ and $\zeta=0.1$.}
\label{fig:partnumb}
\end{figure}

For this purpose, we include a loss term to the GP equation that in this case takes the form
\begin{equation}
i\partial_t\psi=H_0\psi+\frac{g_cN}{A}|\psi|^2\psi-i\gamma\left(|E_L(x)|^2+|E_R(x)|^2\right)\psi
\label{eqn:part_loss}
\end{equation}
where $\gamma$ is a phenomenological parameter corresponding to the particle loss rate. In~\fref{fig:partnumb} the effect of different choices of $\gamma$ on the total particle number is shown. Note that the introduced type of loss term takes into account that more particles are lost at higher field intensities than at lower ones. This is due to the fact that in this regions the spontaneous decay rate leading to recoil momenta which kick the particles out of the condensate is larger.

\begin{figure}
\centering
\includegraphics[width=0.5\textwidth]{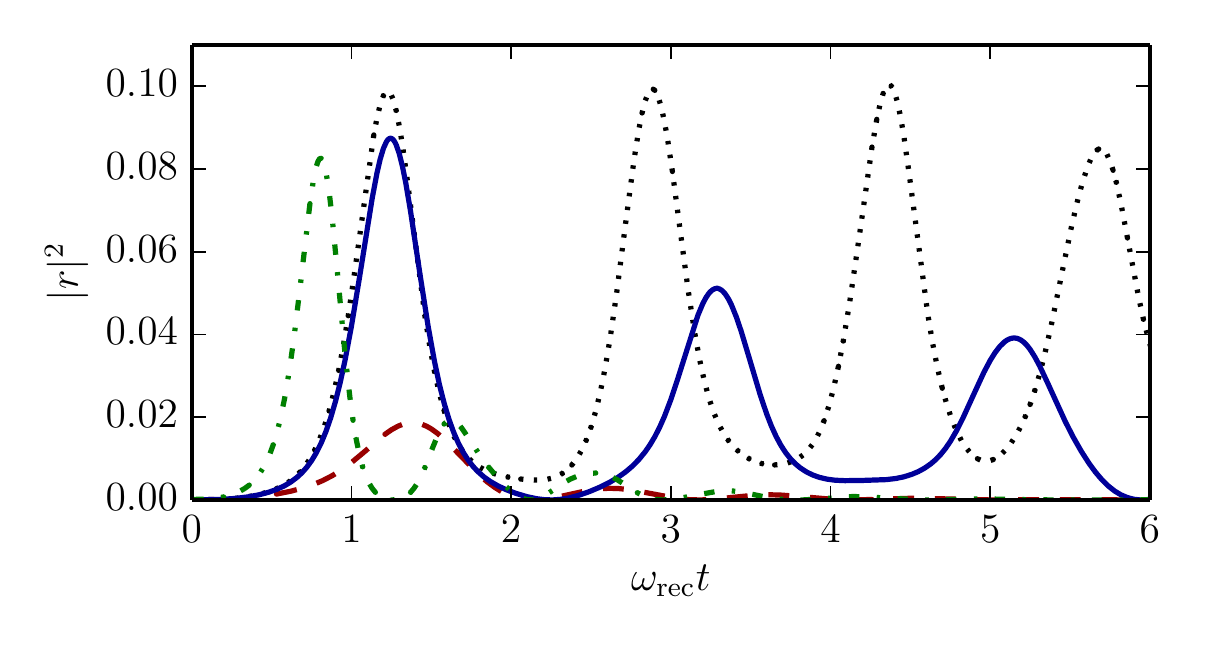}
\caption{Real-time dynamics of the reflection probability for different particle loss rates and particle numbers. $\gamma=0.0$ (dotted black), $\gamma=0.001$ (solid blue) and $\gamma=0.01$ (dashed red). The dash dotted green curve shows the dynamics for a BEC with a particle number which is a factor 1.5 bigger than for the other curves and $\gamma=0.01$. All other parameters as in~\fref{fig:partnumb}.}
\label{fig:part_loss}
\end{figure}

Simulating the real time dynamics of~\eref{eqn:part_loss} and looking at the dynamics of the system's reflectivity shows a clear damping of the dynamics in time. However, the crystallization still takes place as one can see on the exponential growth of $|r|^2$ at short times in~\fref{fig:part_loss}.

Finally, the particle number dependence of the index of refraction is an interesting property as well. For this purpose we fix the loss parameter $\gamma$ and calculate the time evolution of the index of refraction. From the green dash dotted line in~\fref{fig:part_loss} one can see that the dynamics is getting faster for higher particle numbers which is another indication for the collectivity of the effect mentioned above. In addition, it should be mentioned that the particle number dependence of the effect can already be observed in the short time regime without waiting for the crystal to fully form and stabilize. Similar curves as the ones presented in~\fref{fig:part_loss} have also been observed experimentally in a first implementation of the phase transition~\cite{dimitrova2017observation}.

\section{Crystal formation via a slow ramp across the phase transition}\label{sec:ramping}
A routine path to study quantum phase transitions in ultracold gas experiments is an adiabatic ramping of the parameters (fields) between two points in the phase diagram\cite{bloch2008many}. If enough time is given this process is quasi-adiabatic and reversible. As due to losses time was limited in our setup, we introduced a sudden switch of the laser and the resulting quench dynamics of the crystallization process has been studied in detail in~\cite{ostermann2016spontaneous}.
In a quench a lot of interesting physics is going on at the same time, from density fluctuations to phonon excitations of the atom-light crystal. Nevertheless, the process introduces entropy and does not lead to the systems ground state. 

\begin{figure}
\centering
a) \includegraphics[width=0.4\textwidth]{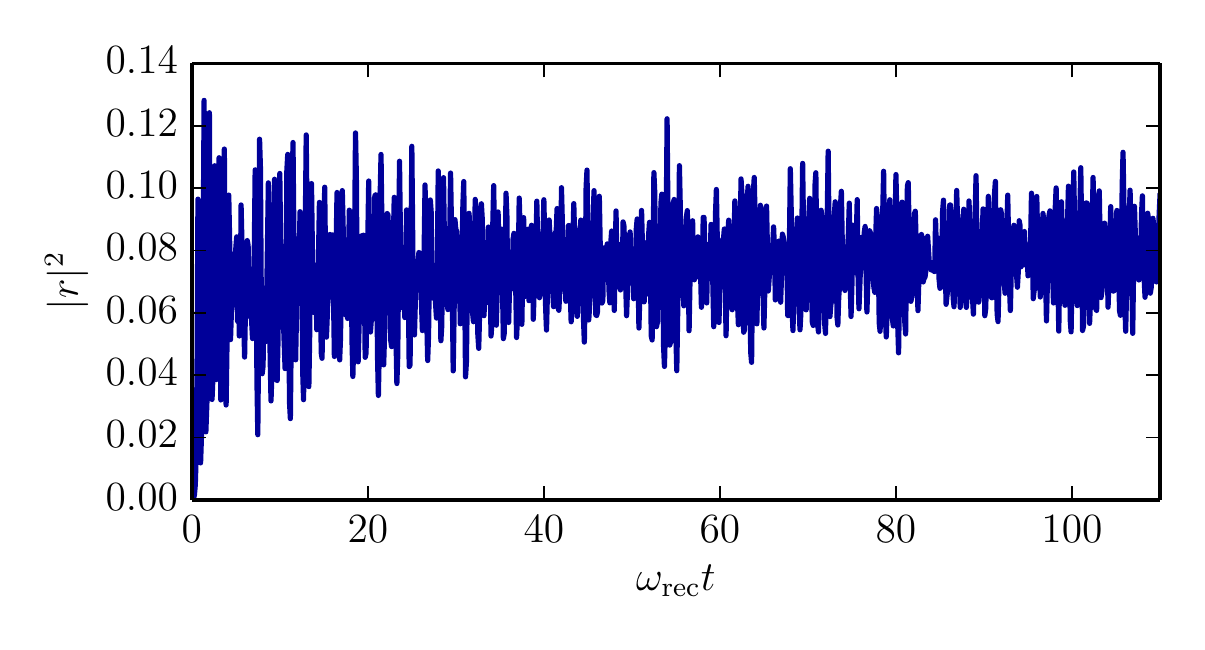}\\
b) \includegraphics[width=0.4\textwidth]{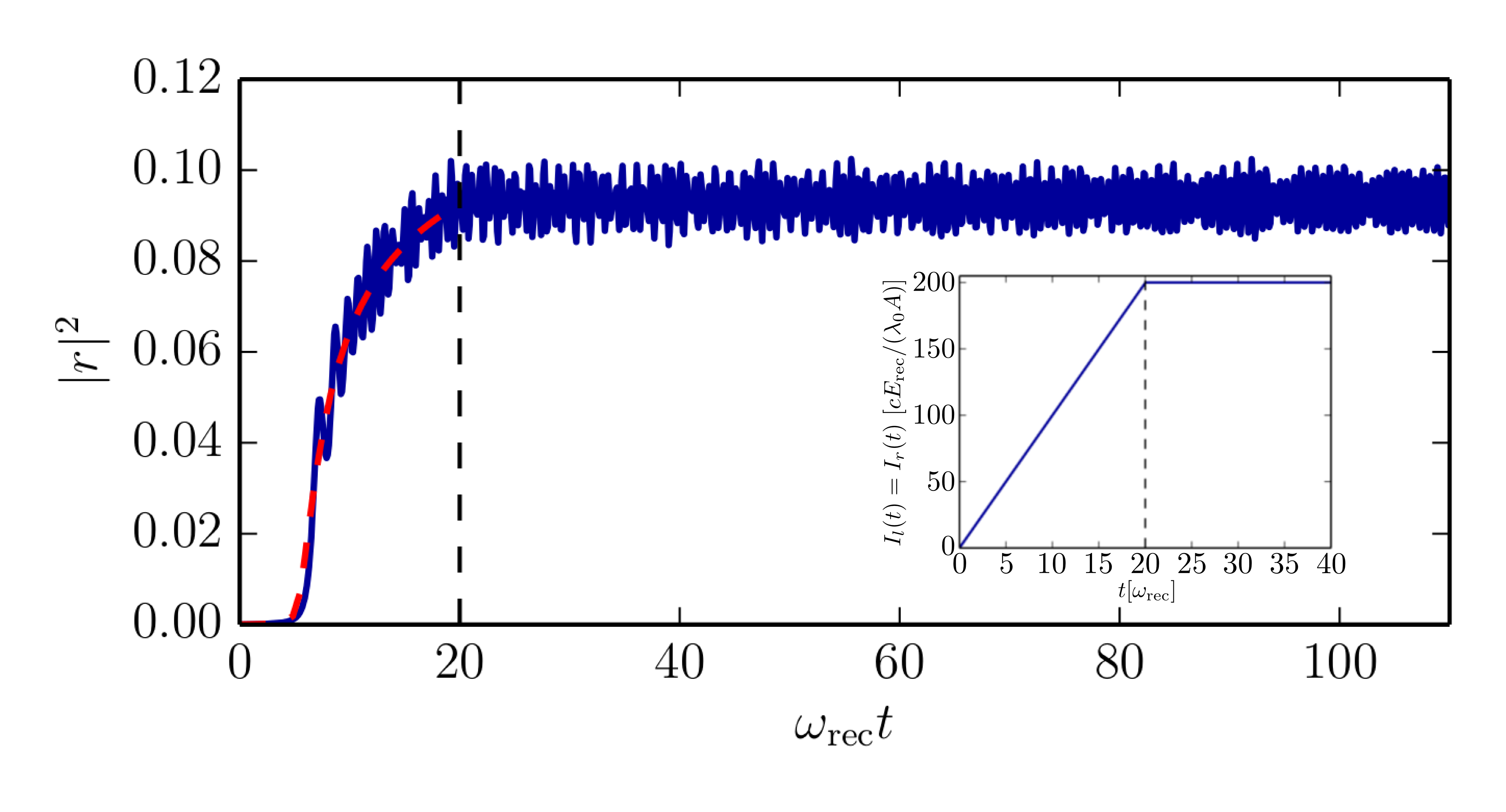}\\
c) \includegraphics[width=0.4\textwidth]{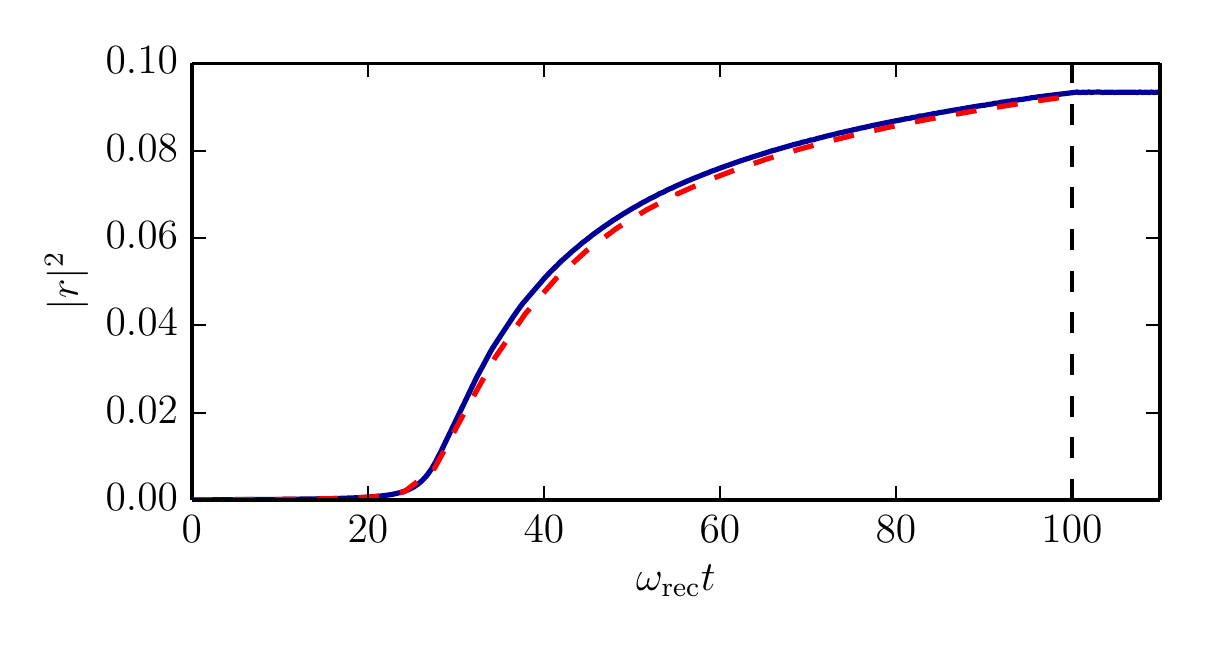}
\caption{Time dependence of total reflection $|r|^2$ for different ramping times a) $t_\mathrm{ramp}=0$, b) $t_\mathrm{ramp}=20$ (The inset figure shows an exemplary ramping function), c) $t_\mathrm{ramp}=100$. The blue curve shows the dynamics for real time evolution and the red dashed line the corresponding imaginary time evolution. The vertical dashed black line indicates the ramping time. $I_l=I_r=200I_0$, $\zeta=0.1$, $L=10\lambda_0$, $g_c N=1.0E_{rec}$.}
\label{fig:ramping_r}
\end{figure}

As an alternative it is thus of specific interest, under which conditions the system's stationary ground state can be reached by performing a sufficiently slow adiabatic passage. For this, one has to ensure that no higher motional states are excited when crossing the threshold, which implies sufficiently large ramping times for
 the system to remain in its ground state during the whole crystallization procedure. Actually as some of our excitations are gapped due to the long range interactions this might be less critical than expected. To determine suitable ramping times the real time evolution for simple linear intensity ramps is numerically evaluated. Please note that no damping is taken into account in this case. The corresponding results are presented in~\fref{fig:ramping_r}. Most likely better ramping functions exist and one could even think of optimal control to improve the results.  

The reflection coefficients again give valuable real time information on the dynamics and one sees how different ramping times affect the dynamics. We see that for ramping times more than twenty times the inverse recoil frequency only little noise is left and an almost perfect adiabatic passage can be realized (see~\fref{fig:ramping_r}).
\begin{figure}
\centering
\includegraphics[width=0.4\textwidth]{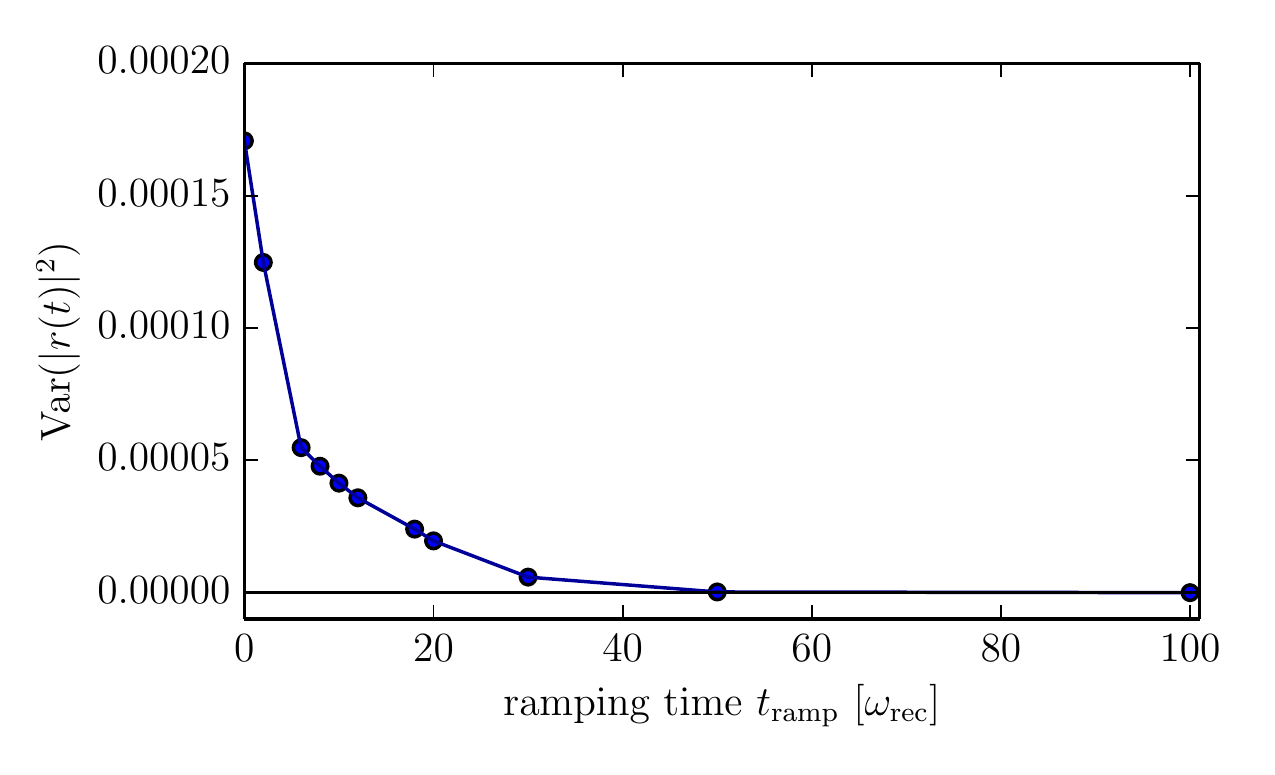}
\caption{Time averaged variance of the time dependent reflection probability for a real-time evolution simulation in the stationary lattice regime during last 50 time steps for different ramping times as a quantitative measure of adiabaticity. The parameters are the same as in~\fref{fig:ramping_r}}
\label{fig:r_variance}
\end{figure}
To quantify the quality of the adiabaticity and the number of phonon like excitations left, we calculate the variance of the reflectivity around it's mean value for the last 50 time steps and plot it as a function of the ramping time. This variance tends to be close to zero for ramping times $t_\mathrm{ramp}>60$ $\omega_\mathrm{rec}$ (see~\fref{fig:r_variance}.)

\begin{figure}
\centering
a) \includegraphics[width=0.4\textwidth]{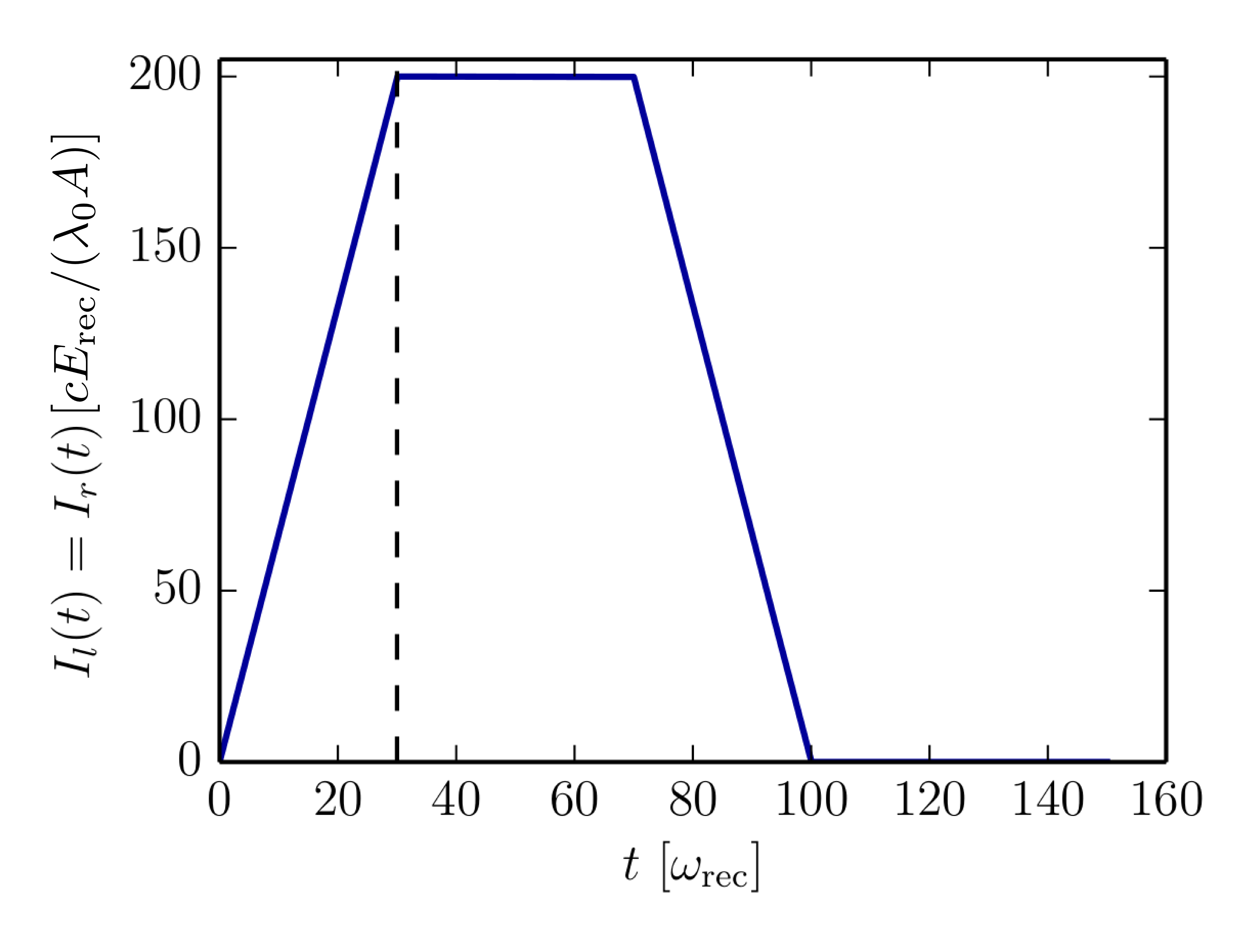}
b) \includegraphics[width=0.41\textwidth]{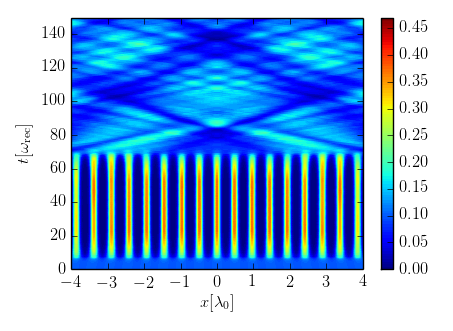}
\caption{a) Ramping function used for quasi adiabatic transfer to the crystalline phase and back b) Real Time evolution of the particle density during a ramp with the same parameters as in~\fref{fig:ramping_r}).}
\label{fig:ramping_both_dir}
\end{figure}

\begin{figure}
\centering
\includegraphics[width=0.45\textwidth]{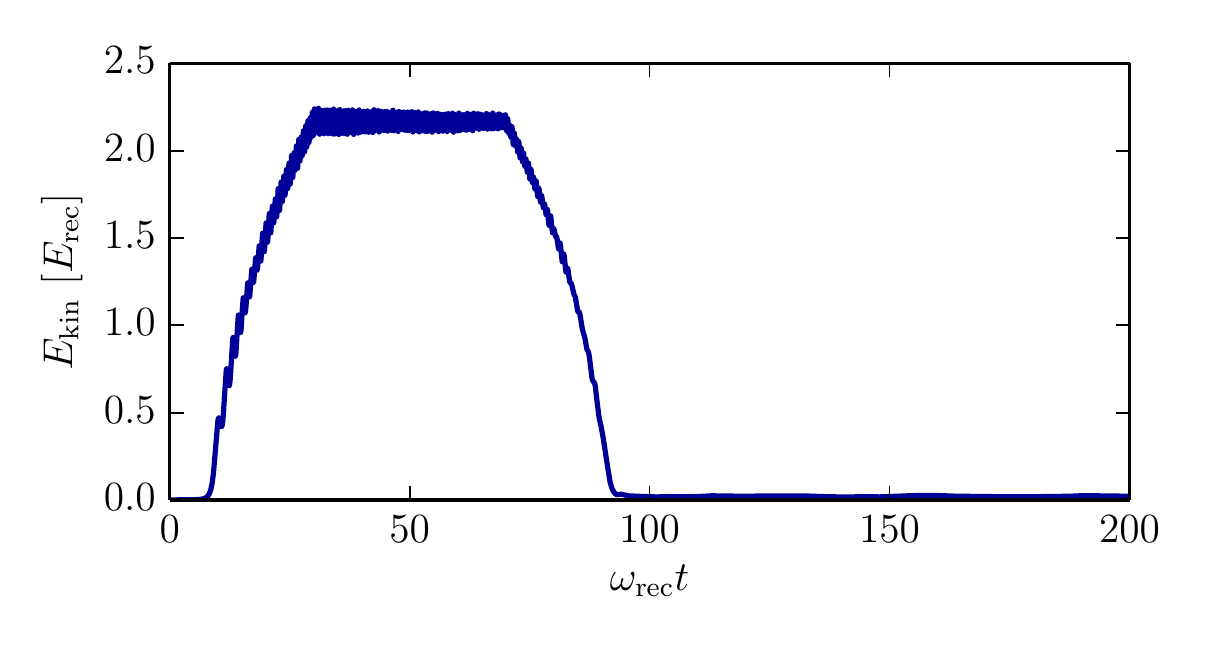}
\caption{Time evolution of the system's kinetic energy during the ramping procedure shown in~\fref{fig:ramping_both_dir}a.}
\label{fig:Ekin}
\end{figure}

Another measure of the introduced heating and entropy through the ramp can be derived from the reversibility of the ramping process without any cooling invoked. For this purpose, one has to look at a ramp realizing a transition from homogeneous order to crystalline order and back (see~\fref{fig:ramping_both_dir}a)). For an ideal ramping procedure one should end back in a perfect BEC in the end. In~\fref{fig:ramping_both_dir}b) the space-time density plot for $|\psi(x,t)|^2$ is shown. The BEC performs the transition from homogeneous to periodic order as expected.
After ramping down the laser intensity again one can recognize that the BEC is not perfectly homogeneous but some density and phase fluctuations are left. This could come as simple fluctuations or even lead to formation of vortex pairs\cite{piazza2015self}.  Nevertheless, the periodic order is completely lost and only some higher excitations remain. It can be claimed that for increasing ramping times this higher excitations will get smaller and smaller. To quantify the energy remaining in the system we calculate at the dynamics of the kinetic energy
\begin{equation}
E_\mathrm{kin}=\frac{1}{2m}\int dx \hbar^2|\partial_x\psi|^2.
\end{equation}
The result is shown in~\fref{fig:Ekin}.
Of course, the kinetic energy initially follows the ramping procedure. The important question is how small it is in presence of higher excitations after the ramping process. Figure~\ref{fig:Ekin} shows that for $t>100\omega_\mathrm{rec}$ the kinetic energy is very small (but not zero). This implies that the heating imposed by the ramping process is small and that despite the small excitations the BEC can be restored. This is very similar to the physics observed in the superfluid to Mott insulator transition in a BEC. In summary, it can be posed that the spontaneous crystallization process can be reversed if the ramping times are sufficiently large.


\section{Crystallization versus matter-wave superradiance}\label{sec:superradiance}

Laser induced instabilities of Bose-Einstein condensates have been found and studied for more than a decades now, starting with seminal experiments at MIT~\cite{inouye1999superradiant}. Already  soon after the early experiments with BECs it became obvious that not only the Bose enhancement from the BEC but also stimulated light emission interpreted as self-stimulated Kapitza-Dirac scattering, was a key mechanism behind these observations ~\cite{schneble2003onset, piovella2001quantum}. In these experiments, the BEC was driven from a single direction and above a certain pump threshold an instability towards a density modulation was found, which in turn enhanced the light scattering in analogy to collective atomic recoil lasing CARL~\cite{piovella2001quantum}. This effect is known as Superradiant Rayleigh scattering (SRS). As single side pumping breaks mirror symmetry and momentum is injected only from one side, obviously no stable final state could exist. Using cavity enhancement allows better control of these scattering processes~\cite{kessler2014steering} and one can even use them for cooling~\cite{sandner2015self}. This directly connects to cavity based selfordering phenomena for standing wave pumping~\cite{ritsch2013cold}.

In our configuration of symmetric counter-propagating pumping along the longitudinal BEC direction, mirror and translational symmetry is kept, but the fundamental microscopic light matter interaction parameters are rather similar.  It is therefore natural to ask what is the relation between spontaneous crystallization, matterwave superradiance and collective atomic recoil lasing. In particular the initial phase of the self-ordering process, where all fields act independently should bear some similarities.  

Apart from the obvious fact that a one-sided setup cannot lead to a steady-state, there seem to be deeper differences between the two phenomena. In fact, spontaneous crystallization can be understood using only optical potentials and dipole forces. Apart from creating initial fluctuations no spontaneous scattering is needed. Both, the light field and the atoms are dynamic mean fields, which exchange energy and momentum in the dispersive regime. Indeed, even though the crystallization threshold~\eref{eqn:Icrit} is proportional to the Rayleigh scattering rate scaling with $1/\delta^2$ as it is the case for matter wave super-radiance, the full dynamics of crystallization actually depends on the effective dipole-potential depth proportional to $1/\delta$. The quadratic dependence of $I_c$ on $\delta$ is only due to the non-linearity of the model, as explained in section~\ref{sec:crystallization}.

\begin{figure}
\centering
a) \includegraphics[width=0.35\textwidth]{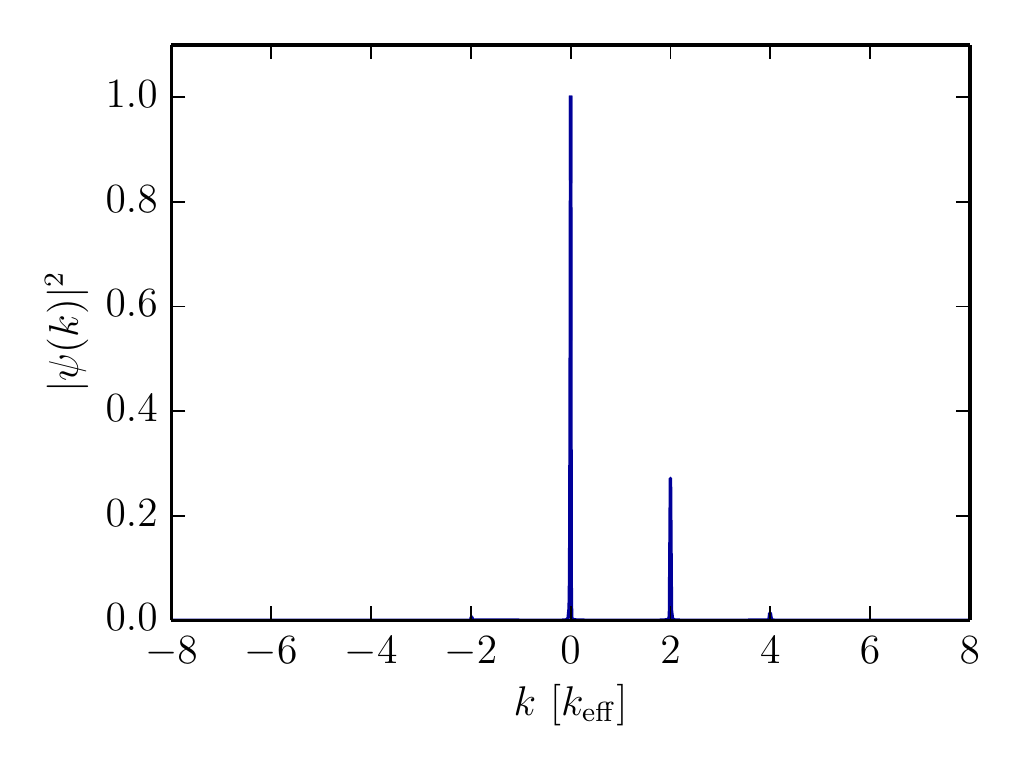}\hspace{1cm}
c) \includegraphics[width=0.35\textwidth]{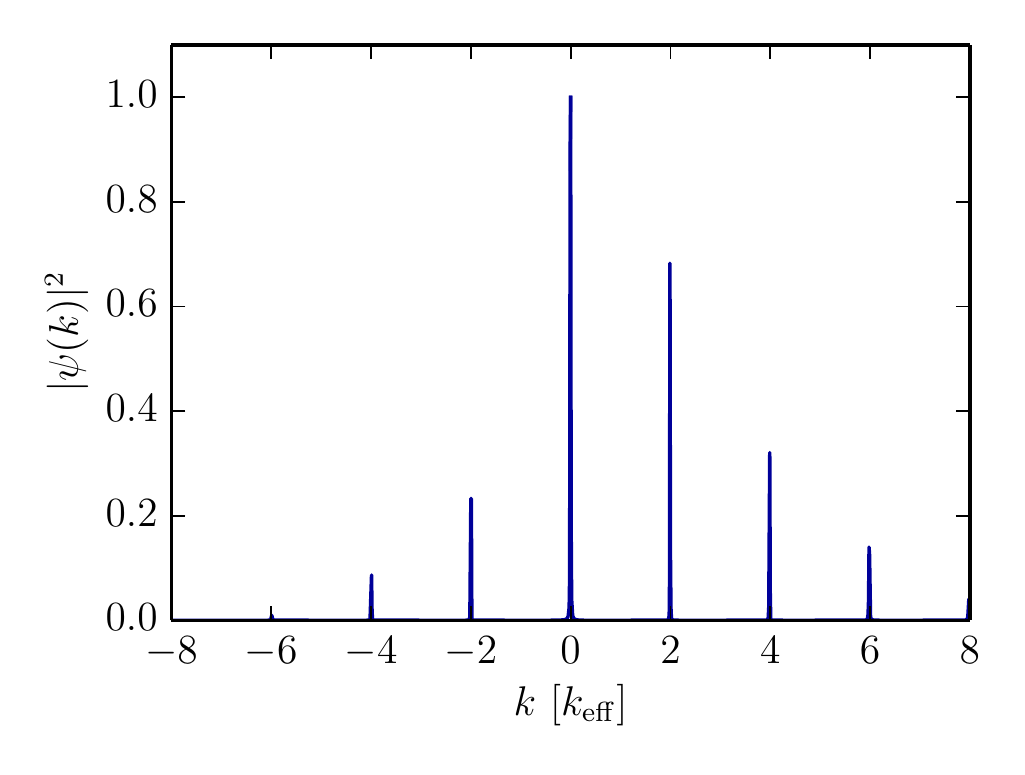}\\
b)\includegraphics[width=0.35\textwidth]{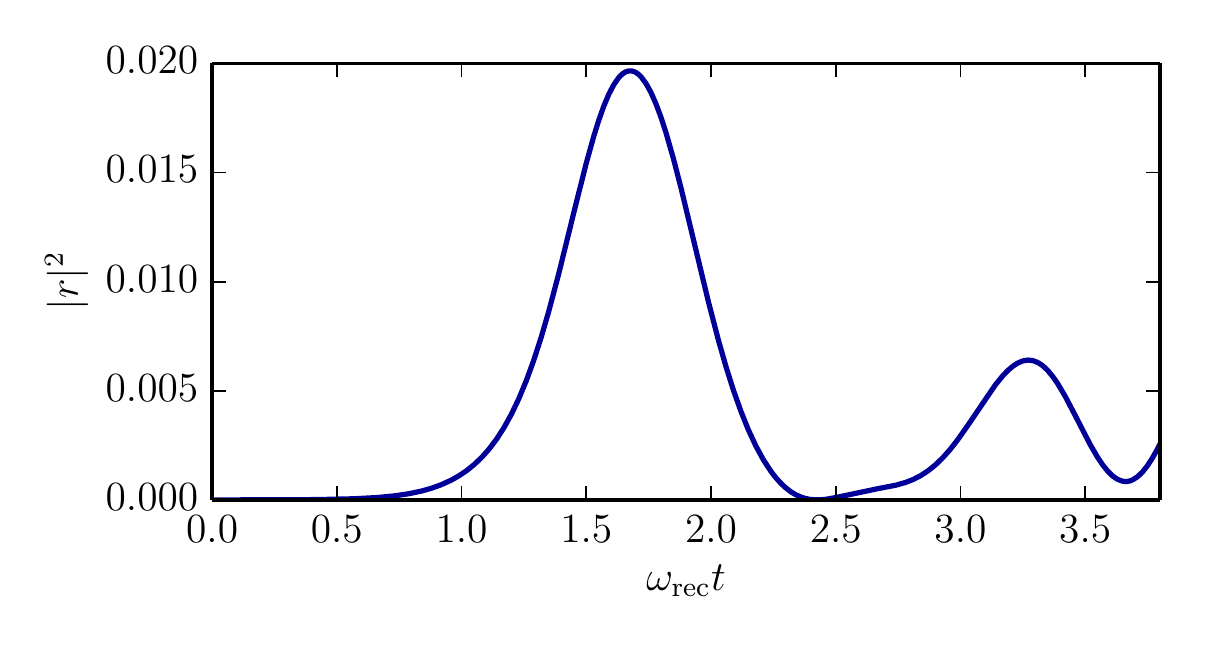}\hspace{1cm}
d) \includegraphics[width=0.35\textwidth]{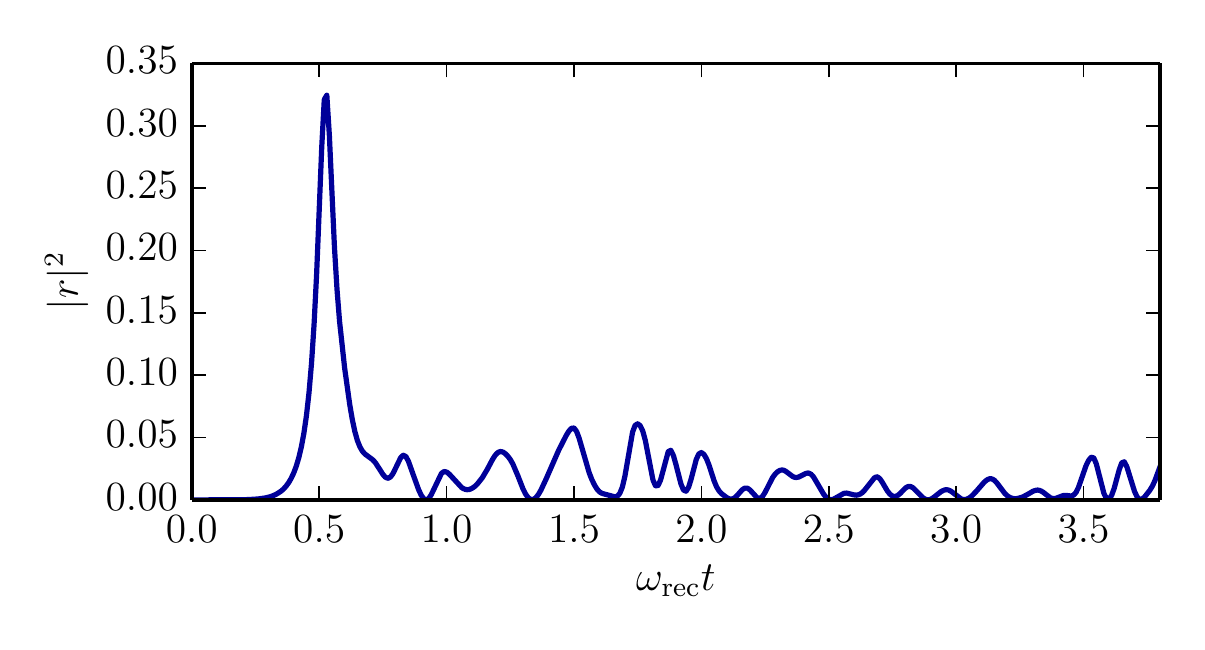}
\caption{a) BEC momentum distribution at  $\omega_\mathrm{rec} t=1.6 $ for $\zeta=0.1$, $g_cN=1$, $L=100\lambda_0$ and $I_l=200$, $I_r=0$ and b) corresponding time dependent reflection coefficient. c) Same as above at $t \omega_\mathrm{rec}=0.4$ for $\zeta=0.2$, $g_cN=1$, $L=100\lambda_0$ and $I_l=300$, $I_r=0$ with d) the evolution of the reflection coefficient.}
\label{fig:superradiance}
\end{figure}

Nevertheless, when applied to the case of single-side driving also the dispersive model shows a modulation instability qualitatively similar to the one observed in~\cite{inouye1999superradiant}.  In~\fref{fig:superradiance} two representative examples are shown.

\begin{figure}
\centering
\includegraphics[width=0.5\textwidth]{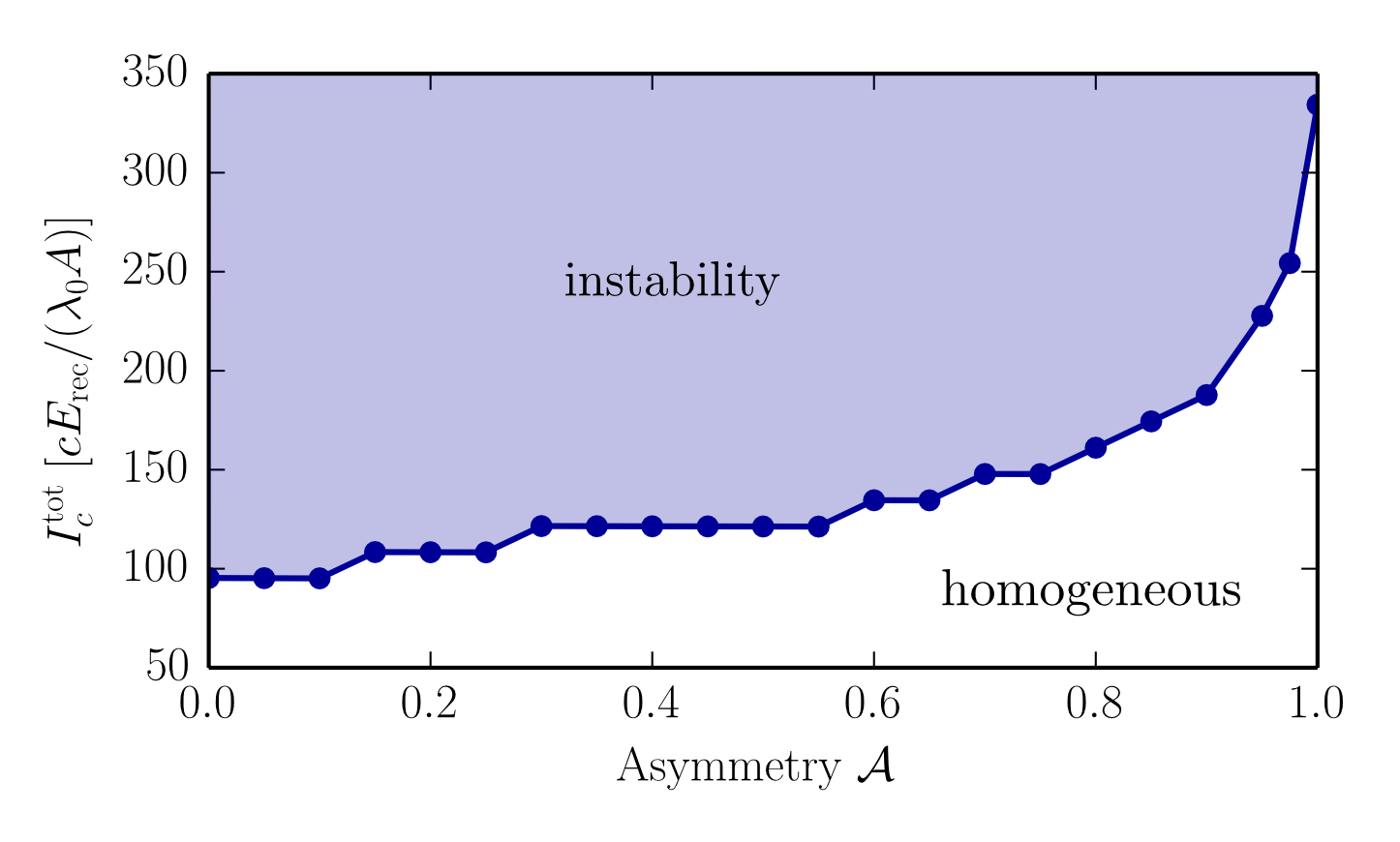}
\caption{Stability region of the homogeneous phase showing the maximal threshold of the total laser power applied from both sides for different pump asymmetries as defined in~\eref{eqn:asym}. The threshold is obtained from calculating the stationary state via imaginary time evolution of the coupled field equations for different total intensities. $\zeta=0.1$, $L=100 \lambda_0$, $g_c N = 1.0 E_{rec}.$}
\label{fig:phasediagramm}
\end{figure}

One clearly recognizes a certain directionality in the Bragg-diffraction pattern, but in general both positive- and negative-momentum peaks are present. In addition, a high initial peak of the reflection coefficient in time shows that at the point where the atomic bunching occurs a light flash can be observed. The effect is getting more prominent if the atom-light interaction strength (\ie the optical density of the medium) is high. 
We also note that the pump threshold for the instability differs from the one for symmetric driving given in Eq.~\eref{eqn:Icrit} only by a geometrical pre-factor of order one. The corresponding phase diagram is shown in~\fref{fig:phasediagramm} where we defined the beam asymmetry
\begin{equation}
\mathcal{A}:=\frac{I_l-I_r}{I_l+I_r}.
\label{eqn:asym}
\end{equation}

Even though the momentum distribution of the gas and the amount of reflected light behave qualitatively similar to the SRS experiment of~\cite{inouye1999superradiant}, there is an important quantitative difference which is an evidence for the different physical origin. The time scale over which the instability appears in our model is of the order of one inverse recoil frequency $1/\omega_{\rm rec}$ or shorter, while in the SRS experiment of~\cite{inouye1999superradiant} the instability grows on time scale much longer than inverse recoil. The difference in the nature of the instability between long and short time scales was experimentally investigated in~\cite{schneble2003onset}. A more recent experiment~\cite{mueller2016semi} showing qualitatively similar results as the one presented in this work and shown in~\fref{fig:superradiance} has been performed in the short-pulse regime.
The comparison of our prediction with the experiments of~\cite{schneble2003onset} and~\cite{mueller2016semi} suggests that the model described in~\ref{sec:model} and the physics of superradiant modulational instabilities induced by dispersive forces and not by Rayleigh scattering should be relevant in the short-pulse regime.

\section{Conclusions}\label{sec:conclusion}
Our numerical simulations exhibit that real monitoring the backscattered light as well as time of flight studies of the atomic momentum distribution contain all the essential signatures to show the concept of spontaneous atom-field crystallization introduced in~\cite{ostermann2016spontaneous}.  While the initial growth of the momentum distributions, apart from a threshold shift, strongly resembles single beam matter wave superradiance, the subsequent locking of the two standing waves is a unanimous sign of long range order in this translation and mirror symmetric configuration. Measuring the particle number dependent stationary reflection coefficient clearly can can be traced back to a collective effect. Additional signatures are contained in shifts and oscillations of the atomic center of mass which can be observed also via the time dependence of the atomic momentum distributions.

As our model includes atomic self-interaction, a quasi-stationary crystal will even form by a quench type sudden laser turn on, but to obtain a low entropy ordered phase a slower close to adiabatic ramping over the phase transition is required. We have seen that varying the ramping procedure presented in section 6 gives insight in the entropy generation during the crystallization process and shows that crystallization is reversible even at rather fast ramp speeds. We also see that symmetric pumping reduces the instability threshold as compared to single side super-radiance and there is strong indication for a maximum possible value of of the pump asymmetry, under which an order state is at least metastable.   

Interestingly there appears no obvious size limit on the crystal and for large enough effective particle number almost a hundred percent of the light of both lasers will be reflected without the crystal becoming unstable as e.g. for the transverse pump case\cite{griesser2013light}. Outside the transition region where all the reflection happens one simple gets an ordinary atomic lattice from the atom reflected beams. 

At this point the central experimental challenge is preparation of a BEC with large enough particle numbers. In fact the available observation time until heating sets in is short in general as an estimation in table I of~\cite{asboth2008optomechanical} shows. However, the observation time increases at least linearly with the particle number which leads to the conclusion that for typical atomic configurations about $10^6$ particles would be needed for clear signatures of stable order. When the system size is further increased by one or two orders of magnitude the formation of a stable crystal with signatures of a gapped phonon spectrum should be possible.

\ack
We would like to thank F. Meinert  as well as W. Lunden, I. Dimitrova and W. Ketterle  for very fruitful discussions on possible experimental implementations and observable signatures presented in the manuscript. We acknowledge support by the Austrian Science Fund FWF through projects SFB FoQuS P13 and I1697-N27.


\end{document}